\documentclass[pdflatex,sn-nature]{sn-jnl}

\usepackage{graphicx}%
\usepackage{multirow}%
\usepackage{amsmath,amssymb,amsfonts}%
\usepackage{amsthm}%
\usepackage{mathrsfs}%
\usepackage[title]{appendix}%
\usepackage{xcolor}%
\usepackage{textcomp}%
\usepackage{manyfoot}%
\usepackage{booktabs}%
\usepackage{algorithm}%
\usepackage{algorithmicx}%
\usepackage{algpseudocode}%
\usepackage{listings}%
\usepackage{comment}
\usepackage{setspace}
\usepackage{pdfpages}

\fontsize{16pt}{24pt}\selectfont
\theoremstyle{thmstyleone}%
\theoremstyle{thmstyletwo}%
\theoremstyle{thmstylethree}%

\begin{document}

\title[Article Title]{Nonuniform Screening Reshapes Collective Excitons in Molecular Aggregates from Stochastic Bethe-Salpeter Theory}

\author*[1]{\fnm{Barry Y.} \sur{Li}}\email{yangtao.li@weizmann.ac.il}
\author*[2]{\fnm{Chern} \sur{Chuang}}\email{chern.chuang@unlv.edu}
\author[3]{\fnm{Tim} \sur{Duong}}\email{td0303@princeton.edu}
\author[4]{\fnm{Tucker} \sur{Allen}}\email{tuckerallen27@g.ucla.edu}
\author[4]{\fnm{Justin R.} \sur{Caram}}\email{jcaram@chem.ucla.edu}
\author[4]{\fnm{Daniel} \sur{Neuhauser}}\email{dxn@ucla.edu}

\affil[1]{\orgdiv{Department of Chemical and Biological Physics}, \orgname{Weizmann Institute of Science}, \orgaddress{\city{Rehovot}, \postcode{7610001}, \country{Israel}}}
\affil[2]{\orgdiv{Department of Chemistry and Biochemistry}, \orgname{University of Nevada, Las Vegas}, \orgaddress{\city{Las Vegas}, \state{NV}, \postcode{89154}, \country{USA}}}
\affil[3]{\orgdiv{Department of Chemistry}, \orgname{Princeton University}, \orgaddress{\city{Princeton}, \state{NJ}, \postcode{08544}, \country{USA}}}
\affil[4]{\orgdiv{Department of Chemistry and Biochemistry}, \orgname{University of California, Los Angeles}, \orgaddress{\city{Los Angeles}, \state{CA}, \postcode{90095}, \country{USA}}}

\abstract{We build and apply an \textit{ab initio} stochastic Bethe-Salpeter Equation (sBSE) approach to planar cyanine dye aggregate clusters containing up to 2,856 valence electrons. Using a transferable, parameterized screened exchange kernel, we resolve dielectric response, exciton delocalization, and optical spectra within a unified many-body framework. Remarkably, the sBSE transition densities validate the Frenkel exciton picture in real space and provide a practical first-principles route to its screened couplings. The calculations reproduce \textit{H}-, \textit{I}-, and \textit{J}-like spectral evolution and reveal that spatially dependent screening does not merely shift excitonic bands, but reshapes their dispersion, leading to changes in the relative energies of bright and dark exciton states. These results establish sBSE as a predictive framework for large molecular aggregates.}


\maketitle

\section{Introduction}\label{sec1}

Polymethine cyanine dye supramolecular assemblies are among the most studied excitonic molecular aggregates due to their chemically modifiable bandgaps and large transition dipole moments, especially within the near- and short-wave infrared spectral windows (NIR: $\lambda=700 - 1000$ nm and SWIR: $\lambda=1000 - 2000$ nm). These aggregates have found applications in biological imaging, lasing, and optoelectronics while also serving as model systems to explore strong dipole coupling and coherent exciton transport.\cite{Scholes2003,Scholes2006,Cheng2009,May2011,Hsu2011,Ringstrom2026,Deshmukh2019,Deshmukh2022,Bailey2023,Williams2024,Bazile2026,Wright2026,Cosco2021,Carr2018} Conventional exciton theory relates aggregate packing to optical response, distinguishing \textit{H}- from \textit{J}-aggregates through the sign of intermolecular exciton coupling. More recent work has shown that planar cyanine aggregates exhibit richer behavior, including intermediate \textit{I}-aggregate characteristics that cannot be described within the traditional \textit{H}/\textit{J} classification.\cite{Jelley1936,Kasha1963,Kasha1965,Davydov1964,Fidder1991,Wurthner2011,Deshmukh2019,Deshmukh2022} For example, in extended two-dimensional aggregates, an \textit{I}-aggregate regime can arise when short- and long-range excitonic couplings have opposite signs, producing a red-shifted bright state that lies within the excitonic band rather than at its lower edge.\cite{Deshmukh2019,Deshmukh2022} Complementary extensions have further shown that competition between short- and long-range excitonic interactions can generate \textit{HH}, \textit{HJ}, \textit{JH}, and \textit{JJ} regimes with distinct excitonic band structures.\cite{Hestand2017,Hestand2018} At the same time, advances in cryogenic electron microscopy have provided structural information with near-atomic resolution, creating a growing need for electronic structure methods that can directly connect aggregate geometry and optical properties.\cite{Deshmukh2024,Jelley1936,Kasha1963,Kasha1965,Davydov1964}

Most theoretical studies of cyanine aggregates use Frenkel exciton Hamiltonians to connect aggregate geometry with photophysical observables.\cite{Davydov1964,Wurthner2011,Fidder1991,Spano2010,Spano2014,Hestand2017,Hestand2018,Chuang2019,Bradbury2020} In these studies, localized molecular excitations are coupled through site energies and intermolecular interactions, enabling efficient treatment of disorder, packing geometry, and large aggregate sizes.\cite{Kuhn1970,Spano2010,Spano2014,Hestand2017,Hestand2018} The predictive accuracy of the Frenkel framework therefore depends on how its site energies, couplings, and environmental screening are constructed. Couplings are commonly approximated using point-dipole, extended-dipole, or transition-charge models, whose accuracy can become sensitive to short-range geometry,\cite{Krueger1998,Madjet2006,Fujimoto2010,Tempelaar2014} while dielectric screening is typically introduced through effective dielectric constants, continuum models, or other parametrized treatments rather than derived directly from the many-body response.\cite{Novoderezhkin2004,Mennucci2013,Megow2015,Curutchet2017}

An \textit{ab initio} approach to treating excitations in extended molecular aggregates would allow for first-principle study of aggregate spectral properties. To that end, time-dependent density-functional theory (TDDFT) is a routine approach for describing molecular aggregate excited states, however, its accuracy relies heavily on the choice of exchange-correlation functional. In turn, the Bethe-Salpeter Equation (BSE) of many-body perturbation theory has become a promising alternative to TDDFT with the ability to describe charge-transfer excitations more reliably. The success of the BSE comes from a screened Coulomb interaction that explicitly accounts for interactions between electrons and holes.\cite{Hedin1965,Strinati1988,Rohlfing2000,Onida2002,Blase2018,Blase2020,Duchemin2018} The BSE has become a benchmark approach for molecular and condensed-phase spectroscopy, but generating the screened Coulomb matrix elements scales quartically with system size. This has limited application to large-scale molecular aggregate settings.\cite{Onida2002,Blase2011,Blase2018,Blase2020,Duchemin2016,Duchemin2018,Jacquemin2015a,Jacquemin2015b,Loos2020,Rebolini2016} As a result, a large gap persists between experimentally relevant molecular aggregates and the many-body methods needed to describe them.

Our recent developments in stochastic many-body theory provide a route toward closing this gap.\cite{Baer2004,Neuhauser2013,Neuhauser2014,Neuhauser2017,Vlcek2017,Rabani2015,Bradbury2023a,Chen2025} By replacing the explicit construction of the real-space screened Coulomb operator with stochastic realizations of the operator's action, these approaches retain the essential many-body physics of the BSE while substantially reducing computational cost. Previous work has shown that stochastic screened-interaction methods reproduce BSE-quality excitation energies for challenging polymethine cyanine dyes while extending excited-state calculations to systems containing several thousand valence electrons.\cite{Bradbury2023b,Bradbury2024,Li2025,Allen2026} In this work, we apply \textit{ab initio} stochastic Bethe-Salpeter Equation (sBSE), also known as TDHF@$v_W$,\cite{Bradbury2023b,Li2025,Allen2026} to aggregate clusters (containing up to 2,856 valence electrons) of 4$'$-(3,5-dimethylphenyl)-substituted heptamethine cyanine dye (Cy7-DiMePh).\cite{Williams2024,Bailey2023,Deshmukh2019,Deshmukh2020} This NIR-SWIR dye forms multiple well-characterized two-dimensional aggregate mesophases with distinct excitonic and optical properties, while its packing and spectral response are highly sensitive to the 4$'$ aryl substitution.\cite{Williams2024} These features make Cy7-DiMePh a particularly useful model system for connecting molecular packing, dielectric screening, and the emergence of collective exciton structure.\cite{Williams2024}

\begin{figure}[H]
\centering
\includegraphics[width=1\textwidth]{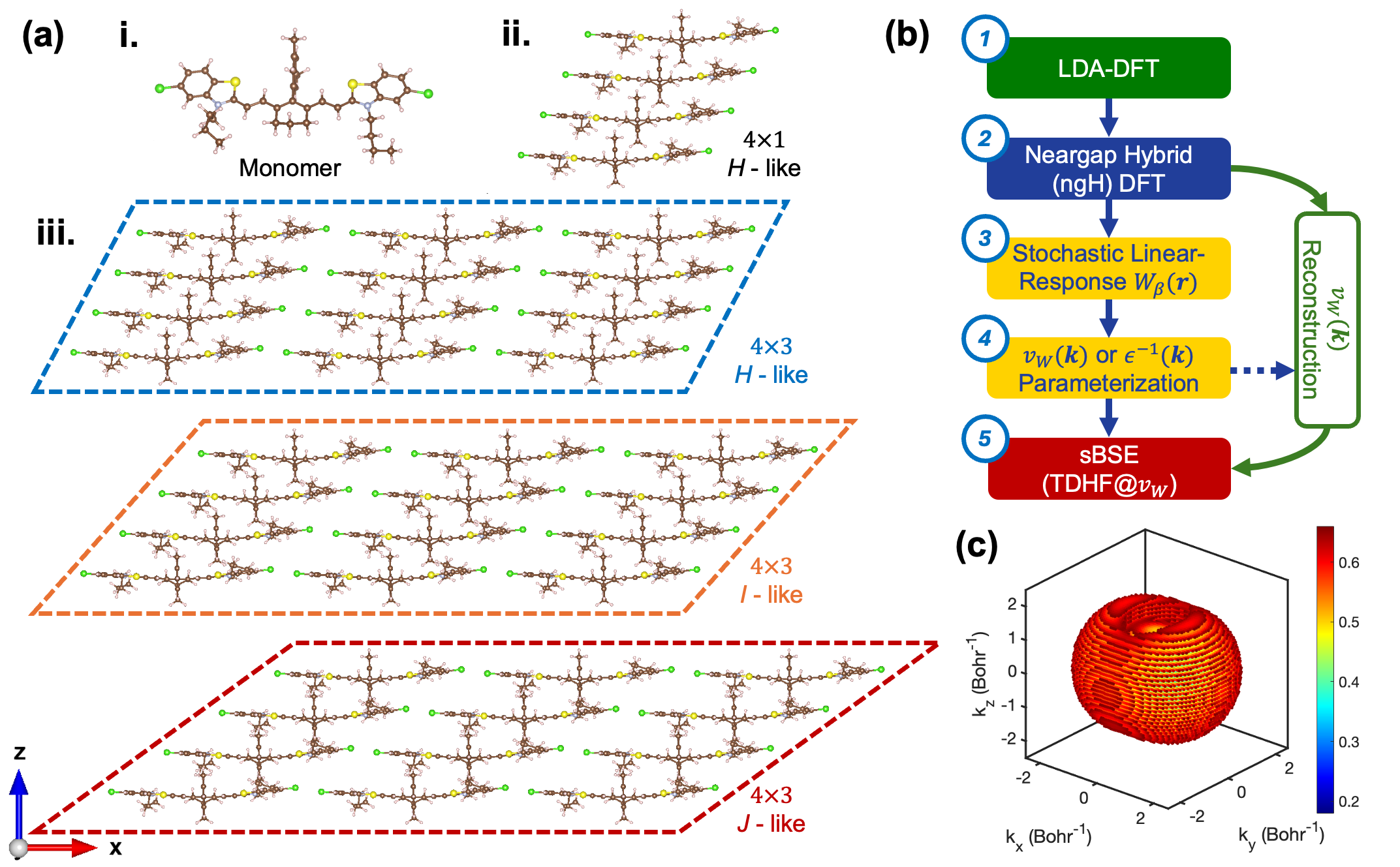}
\caption{\textbf{(a) i.} Cy7-DiMePh monomer structure (terminal sulfonate groups on the side chains are replaced by methyl groups for computational simplicity). \textbf{ii.} 1D $4\times1$ aggregate exhibiting \textit{H}-like stacking motifs along the $\pi$-$\pi$ stacking direction. \textbf{iii.} 2D $4\times3$ aggregates with \textit{H}-, \textit{I}-, and \textit{J}-like packing arrangements. \textbf{(b)} Schematic overview of the sBSE methodology used in this work. \textbf{(c)} Representative reciprocal-space inverse dielectric response, $\epsilon^{-1}(\boldsymbol{k})$, of the Cy7-DiMePh monomer reconstructed from stochastic screening calculations.}\label{fig1}
\end{figure}

Fig.~\ref{fig1} summarizes the systems and workflow considered in this study. Starting from the isolated monomer (Fig.~\ref{fig1} (a) i.), we construct one-dimensional (1D) $2\times1$ (dimer), $3\times1$ (trimer), and $4\times1$ (tetramer) aggregates as well as a series of two-dimensional (2D) $4\times3$ aggregates exhibiting \textit{H}-, \textit{I}-, and \textit{J}-like packing motifs (Fig.~\ref{fig1} (a) ii, iii). The monomer geometry is optimized at the B3LYP/def2-TZVPPD level,\cite{Yanai2004,Weigend2005,Rappoport2010} the $\pi$-$\pi$ stacking distance of $4.72\,\text{\AA}$ is determined by a fragment-constrained dimer geometry optimization at the $r^2$SCAN-3c level,\cite{Grimme2021} and the lateral distance of $25.43\,\text{\AA}$ is determined by a potential energy surface scan using $r^2$SCAN-3c.\cite{Grimme2021} All geometry optimizations are performed using the ORCA 6.0 package.\cite{Neese2012,Neese2025} Additional structural details are provided in the Supplementary Information (SI). The resulting aggregates span a range of experimentally motivated inter-layer slips, with $2.60\,\text{\AA}$, $4.60\,\text{\AA}$, and $6.60\,\text{\AA}$ slips for the \textit{H}-, \textit{I}-, and \textit{J}-like motifs, respectively. Therefore, these aggregates sample distinct intermolecular packing arrangements and excitonic coupling regimes.\cite{Deshmukh2019,Deshmukh2022,Deshmukh2024,Megow2015}

The sBSE workflow is shown in Fig.~\ref{fig1} (b). Ground-state density-functional theory (DFT) calculations are followed by stochastic linear-response time-dependent Hartree calculations that determine the screened interaction. The resulting screening is represented through a polynomial-Gaussian parameterization and incorporated into the sBSE Hamiltonian for excited-state calculations. See the Computational Methods section and SI for more information.

A central quantity throughout this work is the reciprocal-space inverse dielectric function, $\epsilon^{-1}(\boldsymbol{k})$, illustrated in Fig.~\ref{fig1} (c). Unlike conventional dielectric constants, the present framework elucidates the direction- and range-dependent dielectric screening and allows direct examination of how aggregate packing modifies screened electron-hole interactions. We then incorporate this dielectric function into the Frenkel exciton model. Together with the real-space transition-density projection, this provides both a many-body validation of the Frenkel exciton model and a practical scheme for extracting its screened parameters from electronic-structure calculations.

\section{Results and Discussion}\label{sec2}

\subsection{Emergence of \textit{ab initio} dielectrics in molecular aggregates}

\begin{figure}[H]
\centering
\includegraphics[width=1\textwidth]{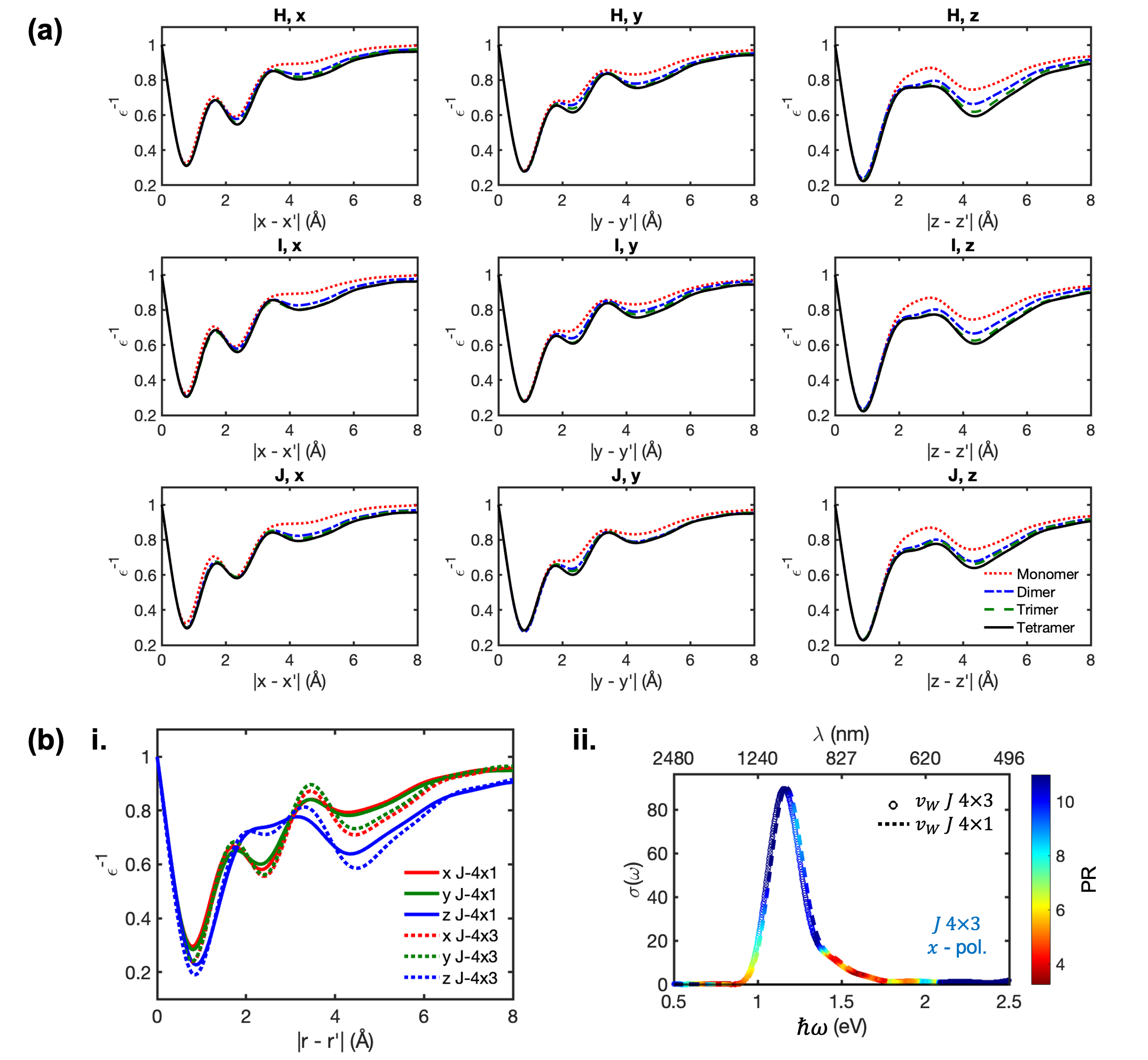}
\caption{\textbf{(a)} Real-space inverse dielectric response, $\epsilon^{-1}(|\boldsymbol{r}-\boldsymbol{r}'|)$, along the $|x-x'|$, $|y-y'|$, and $|z-z'|$ slices for the Cy7-DiMePh monomer and \textit{H}-, \textit{I}-, and \textit{J}-like 1D aggregates. The red, blue, green, and black curves correspond to the monomer, dimer, trimer, and tetramer, respectively. \textbf{(b) i.} Comparison of the $\epsilon^{-1}(|\boldsymbol{r}-\boldsymbol{r}'|)$ slices reconstructed using the $v_W$ obtained from the $4\times1$ and $4\times3$ \textit{J}-like aggregates. \textbf{ii.} $x$-polarized sBSE absorption spectra with PRs for the \textit{J} $4\times3$ aggregate using its own $v_W$ and the \textit{J} $4\times1$ $v_W$.}\label{fig2}
\end{figure}

To understand how intermolecular aggregation modifies the screened electron-hole interaction, we reconstruct the real-space inverse dielectric response, $\epsilon^{-1}(|\boldsymbol{r}-\boldsymbol{r}'|)$, from the reciprocal-space screened interaction kernel, $v_W(\boldsymbol{k})$. Fig.~\ref{fig2} summarizes the resulting $\epsilon^{-1}(|\boldsymbol{r}-\boldsymbol{r}'|)$ for monomer and aggregate Cy7-DiMePh systems.

Fig.~\ref{fig2} (a) shows the real-space Cartesian slices of $\epsilon^{-1}(|\boldsymbol{r}-\boldsymbol{r}'|)$ for monomer, dimer, trimer, and tetramer systems, separated by \textit{H}-, \textit{I}-, and \textit{J}-like packing motifs (across rows) and $x$, $y$, and $z$ directions (across columns). These cuts reveal the full $\boldsymbol{r}$-dependence of the screened interaction, where there is a sharp short-range reduction of $\epsilon^{-1}$ followed by a gradual mid- to long-range recovery toward the unscreened limit of $\epsilon^{-1}\rightarrow1$. This mid- and long-range region is crucial for aggregate excitons, as the electron-hole interaction extends across multiple chromophores. The directional cuts reveal dielectric anisotropy as well. Along the $x$- and $y$-directions, the screening minimum ($\epsilon^{-1}_\text{min}\sim0.28$) occurs at short distances near 0.80 $\text{\AA}$. In contrast, the $z$-direction, which corresponds to the $\pi$-$\pi$ stacking direction, shows a deeper and broader screened region with a later turning point ($\epsilon^{-1}_\text{min}\sim0.23$ around 0.90 $\text{\AA}$), indicating that aggregation builds a more extended dielectric response along the packing axis.

Importantly, the screening profiles converge rapidly with subsystem size. From Fig.~\ref{fig2}  (a), $\epsilon^{-1}(|\boldsymbol{r}-\boldsymbol{r}'|)$ of the dimer, trimer, and tetramer becomes nearly indistinguishable beyond short-range structure, whereas the isolated monomer remains distinct from this converged behavior. This behavior strongly suggests that the screened interaction becomes locally saturated once the immediate dielectric environment surrounding a molecule is adequately represented.

\subsection{Transferability of screened interactions}

This rapid convergence seen in Fig.~\ref{fig2} (a) has important implications for scalable excited-state simulations. Rather than performing the computationally demanding $W_\beta(\boldsymbol{r})$ calculation (defined in the Computational Methods section and the SI) individually for increasingly large aggregates, smaller, locally converged subsystems may provide transferable screened interactions for substantially larger assemblies.

Fig.~\ref{fig2} (b) i. compares $\epsilon^{-1}(|\boldsymbol{r}-\boldsymbol{r}'|)$ slices obtained from the $4\times1$ and $4\times3$ \textit{J}-like aggregates. Despite the increase in total system size, the screening curves remain reasonably similar across all principal directions, especially at long-range. Such similarities indicate that the dominant dielectric environment is primarily determined by local packing geometry. Fig.~\ref{fig2} (b) ii. shows $x$-polarized sBSE spectra with participation ratios (PRs, defined in the SI) of the $4\times3$ \textit{J}-like aggregate computed using the screening kernel, $v_W$, derived from both itself and a smaller $4\times1$ \textit{J}-like subsystem. The two spectra exhibit nearly identical bright peak positions ($<0.02$ eV deviation), side bands, and PRs among excitation energy, $\omega$. Therefore, in the latter simulations, we use the dielectric screening parameterized from the $4\times1$ structures for the bigger $4\times3$ 2D clusters.

\subsection{Aggregate-dependent exciton spectral evolution}

\begin{figure}[H]
\centering
\includegraphics[width=1\textwidth]{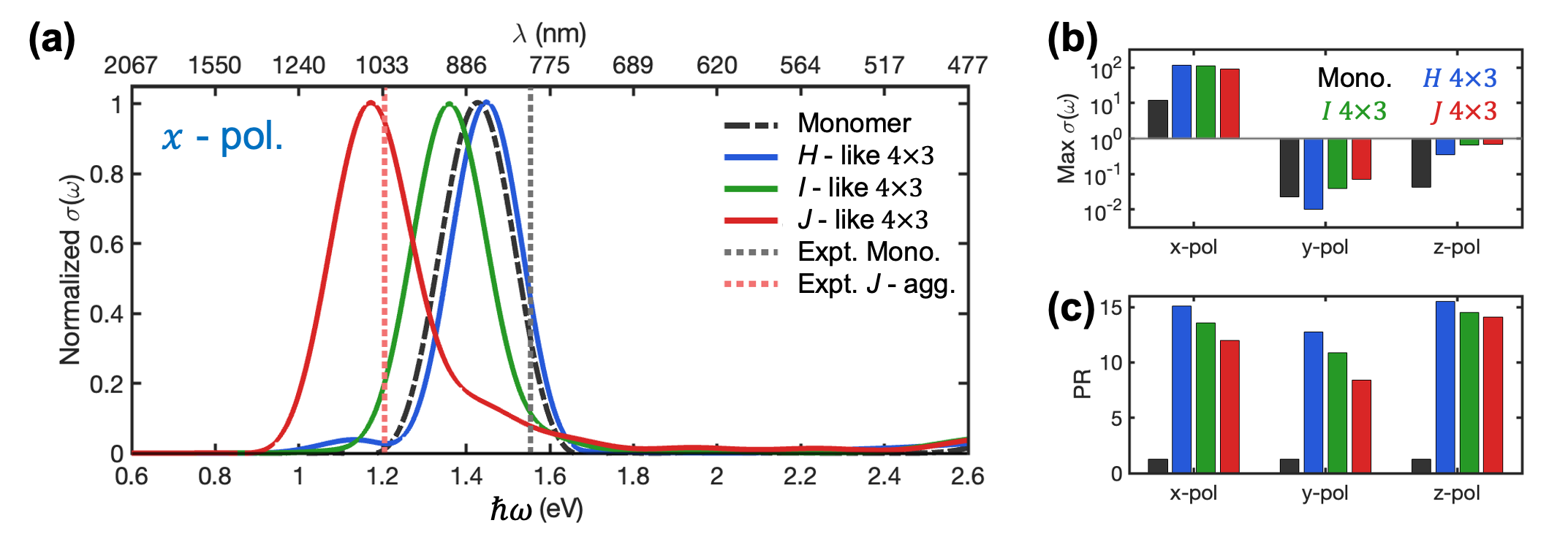}
\caption{\textbf{(a)} Normalized sBSE absorption spectra for the Cy7-DiMePh monomer and $4\times3$ \textit{H}-, \textit{I}-, and \textit{J}-like aggregates under $x$-polarization. Vertical dotted lines indicate the experimental absorption maxima of the monomer at 798 nm (1.55 eV, gray) and the extended \textit{J}-aggregate at 1029 nm (1.20 eV, red).\cite{Williams2024} \textbf{(b)} Maximum absorption cross section, $\sigma(\omega)$, within the $1-2$ eV window for $x$-, $y$-, and $z$-polarized excitations across the monomer and $4\times3$ systems. Cross sections are plotted on a logarithmic scale. \textbf{(c)} PR associated with the brightest exciton within the $1-2$ eV window for different polarizations and aggregate motifs.}\label{fig3}
\end{figure}

Fig.~\ref{fig3} (a) shows normalized $x$-polarized sBSE absorption spectra for the monomer and $4\times3$ aggregates. The monomer has a bright peak at $1.43$ eV ($869$ nm) and $\text{PR}=1.3$. Upon aggregation, the peak shifts according to the packing motif. The \textit{H}-like aggregate blue-shifts to $1.45$~eV ($857$ nm), the \textit{I}-like aggregate red-shifts to $1.36$ eV ($912$ nm), and the \textit{J}-like aggregate red-shifts further to $1.17$ eV ($1057$ nm). In our previous experiments, we measured absorption maxima of $1.55$ eV ($798$ nm) for the Cy7-DiMePh monomer and $1.20$ eV ($1029$ nm) for the extended \textit{J}-aggregate.\cite{Williams2024} Thus, sBSE predicts a monomer-to-\textit{J} redshift of $0.26$ eV, compared with the measured shift of $0.35$ eV. These trends agree with Frenkel exciton and transition-dipole coupling models commonly used to describe cyanine aggregates.\cite{Davydov1964,Spano2010,Hestand2018,Deshmukh2019,Chuang2019,Deshmukh2022,Williams2024} However, unlike simplified dipole-coupling models, the present sBSE calculations incorporate explicit screened electron-hole interactions and many-body dielectric response. Fig.~\ref{fig3} (b) illustrates that the optical responses within the $1-2$ eV window are dominated by $x$-polarization, consistent with the cyanine transition dipole lying along the molecular ``long axis". The peak cross section increases from $12.1$ for the monomer to $\sim10^2$ for the \textit{H}-, \textit{I}-, and \textit{J}-like aggregates. Such enhancement is consistent with the emergence of delocalized exciton behavior expected for coherently coupled cyanine assemblies.\cite{Bailey2023,Williams2024} In contrast, the $y$- and $z$-polarized peak cross sections remain minuscule ($\sim10^{-1}-10^{-2}$).

The PR values in Fig.~\ref{fig3} (c) show that aggregation converts the monomer excitation into a collective exciton. For the brightest $x$-polarized peaks, the PR increases from $1.3$ in the monomer to $15.1$, $13.5$, and $12.0$ for the \textit{H}-, \textit{I}-, and \textit{J}-like aggregates, respectively. The gradual decrease from \textit{H} to \textit{J} indicates that exciton delocalization depends not only on aggregate size but also on packing geometry. The \textit{J}-like aggregate therefore exhibits the largest red shift but not the largest PR, indicating that spectral shift and configurational delocalization are distinct observables. With this, the spectra distinguish two related but different effects: spectral shifts reflect the magnitude and sign of exciton coupling, while the PR quantifies how broadly the excitation is distributed within the electron-hole basis; the validity of the Frenkel representation is assessed separately through the real-space projection described in later sections.

\subsection{Real-space screened transition electrostatic potential}

To visualize the real-space effect of dielectric screening on the exciton transition density, we construct the corresponding transition electrostatic potential using $n_\text{tr}(\boldsymbol{r})$. The bare transition electrostatic potential, $\Phi_\text{tr}^\text{bare}(\boldsymbol{r})$, represents the potential generated by the transition density through the bare Coulomb interaction. Dielectric screening is subsequently incorporated through the $v_W$-derived $\epsilon^{-1}(|\boldsymbol{r}-\boldsymbol{r}'|)$ function. Thus, $\Phi_\text{tr}^\text{scr}(\boldsymbol{r})$ provides a real-space diagnostic of how the screened interaction redistributes this potential relative to the bare Coulomb case.

\begin{figure}[H]
\centering
\includegraphics[width=1\textwidth]{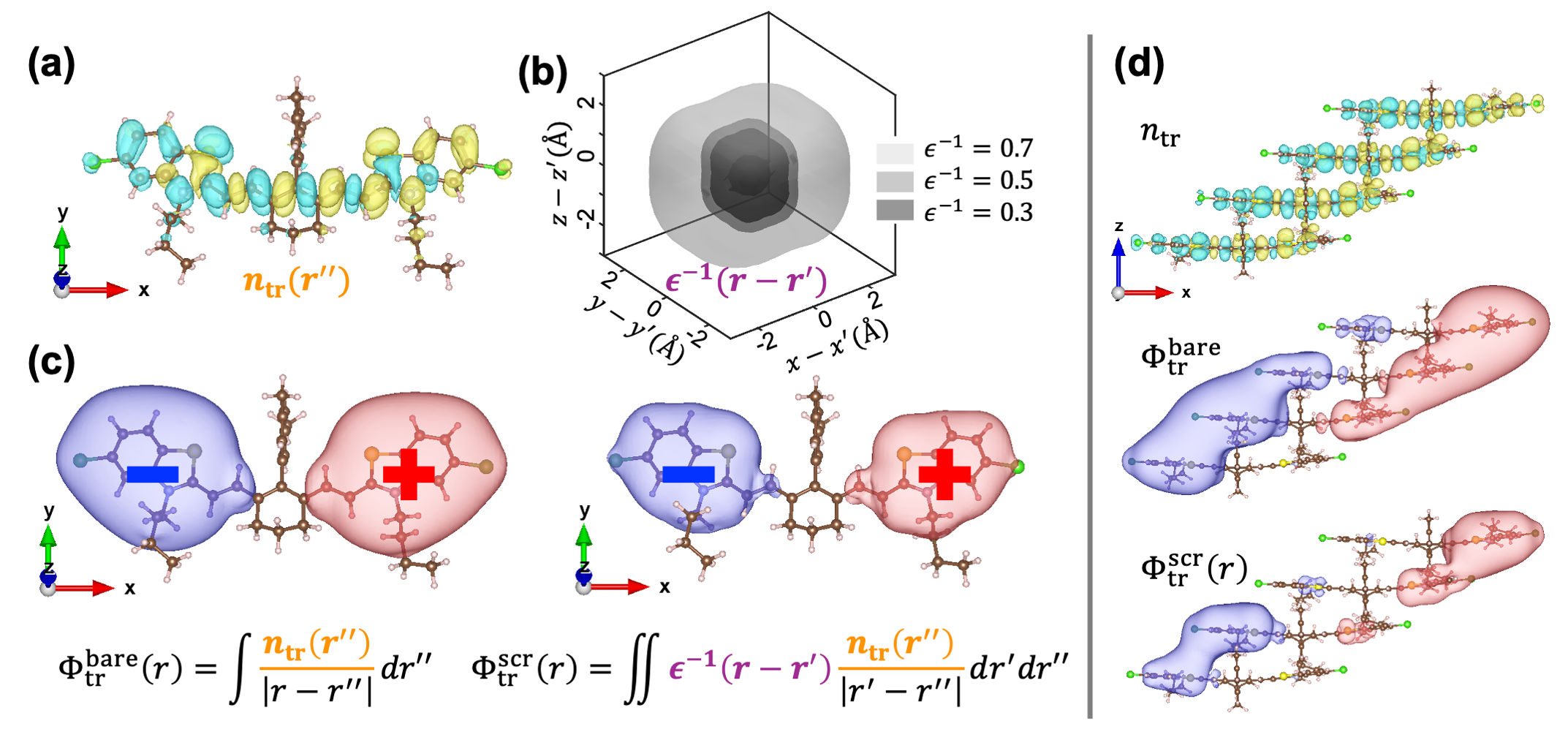}
\caption{\textbf{(a)} Real-space transition density, $n_\text{tr}(\boldsymbol{r})$, of the lowest bright $x$-polarized excitation for the monomer obtained from sBSE. The yellow and cyan isosurfaces represent electron-hole transition amplitudes of opposite sign. \textbf{(b)} Full 3D isosurfaces of $\epsilon^{-1}(|\boldsymbol{r}-\boldsymbol{r}'|)$ for the monomer. Isosurfaces are shown for $\epsilon^{-1}=0.3$, $0.5$, and $0.7$. Additional apparent shades arise from transparency and overlap of the surfaces. \textbf{(c)} Monomer bare and screened transition electrostatic potentials, $\Phi_\text{tr}^\text{bare}(\boldsymbol{r})$ and $\Phi_\text{tr}^\text{scr}(\boldsymbol{r})$, respectively, represented by isosurfaces. \textbf{(d)} $n_\text{tr}(\boldsymbol{r})$, $\Phi_\text{tr}^\text{bare}(\boldsymbol{r})$, and $\Phi_\text{tr}^\text{scr}(\boldsymbol{r})$ for the lowest bright $x$-polarized excitation of the \textit{J} $4\times1$ aggregate. Isosurface values of $\pm0.15\,ea_0^{-3}$ are used for the transition density, while potential isosurface values of $+8$ a.u. (red) and $-8$ a.u. (blue) are used for the bare and screened transition electrostatic potentials of both the monomer and the \textit{J} $4\times1$ aggregate.}\label{fig4}
\end{figure}

Fig.~\ref{fig4} (a)-(c) illustrate the construction of these quantities for the lowest bright $x$-polarized excitation of the monomer. The transition density in Fig.~\ref{fig4} (a) exhibits the familiar alternating positive and negative lobes associated with the electron-hole coherence of the excitation. Fig.~\ref{fig4} (b) shows the reconstructed three-dimensional (3D) real-space screening isosurfaces of the monomer at $\epsilon^{-1}=0.3,\,0.5$, and $0.7$. Integrating $n_\text{tr}(\boldsymbol{r})$ under a Coulomb kernel transforms this rapidly oscillating transition density into a smooth dipolar electrostatic potential composed of spatially separated positive and negative lobes (Fig.~\ref{fig4} (c), left). Incorporating dielectric screening preserves the overall dipolar topology while substantially reducing the spatial extent of the electrostatic potential (Fig.~\ref{fig4} (c), right). Rather than altering the qualitative electrostatic character of the excitation, the screening kernel acts to localize the transition-induced electrostatic potential through the nonlocal dielectric response of the molecular environment. Fig.~\ref{fig4} (d) shows the same ingredients as the monomer, but for a $4\times1$ \textit{J}-like aggregate cluster.  

Together, these results show that dielectric screening primarily modifies the spatial extent of the transition-induced potential while preserving its overall dipolar character.

\subsection{Collective real-space exciton transition in aggregates}
\begin{figure}[H]
\centering
\includegraphics[width=1\textwidth]{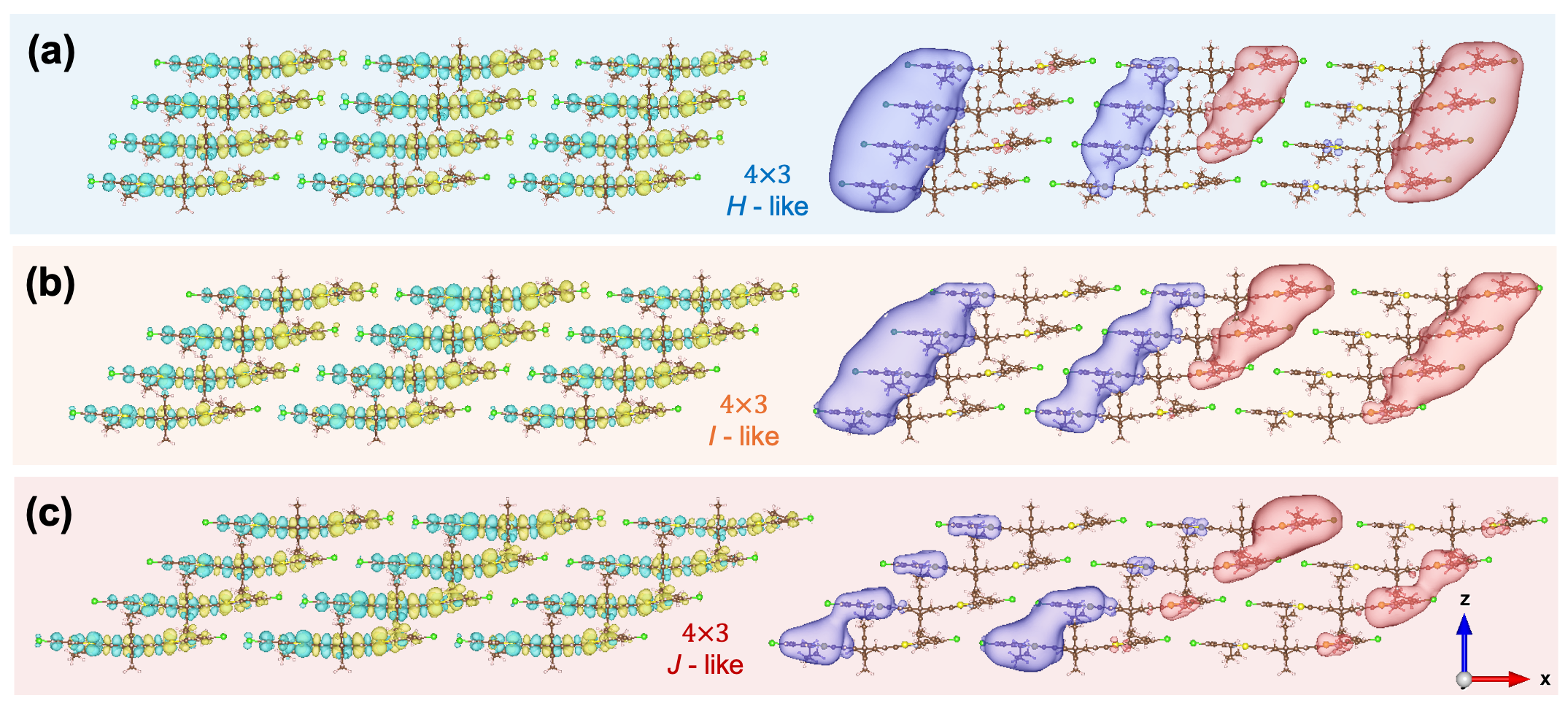}
\caption{Exciton transition density (left) and screened transition electrostatic potential (right) associated with the lowest bright $x$-polarized excitation of \textbf{(a)} \textit{H}-, \textbf{(b)} \textit{I}-, and \textbf{(c)} \textit{J}-like aggregate clusters. The transition density isosurface values of $\pm0.20\,ea_0^{-3}$ and the potential isosurface values of $+10$ a.u. (red) and $-10$ a.u. (blue) are used for all three systems.}\label{fig5}
\end{figure}

To visualize the spatial structure of the collective excitations in 2D aggregation motifs, we compute real-space exciton transition densities for the brightest aggregate excitations together with their transition electrostatic potentials (Fig.~\ref{fig5}). Notably, all the aggregate motifs exhibit spatially extended transition amplitudes aligned predominantly along the aggregate axis. Such delocalized transition density patterns are consistent with the collective exciton states predicted in coherent Frenkel exciton descriptions of cyanine aggregates.\cite{Davydov1964,Spano2010,Hestand2018,Deshmukh2019,Chuang2019,Deshmukh2020,Deshmukh2022,Williams2024} Importantly, the present transition densities emerge directly from the screened many-body sBSE framework rather than from phenomenological exciton Hamiltonian. Therefore, the resulting collective exciton structures naturally incorporate dielectric screening, explicit electron-hole interactions, and packing-dependent electronic structure effects.

While the transition densities remain highly collective and extend over multiple chromophores for all three aggregate clusters (Fig.~\ref{fig5} (a)-(c), left), their corresponding screened transition electrostatic potentials become progressively more compact (Fig.~\ref{fig5} (a)-(c), right), as identical absolute potential isosurface values are used throughout. Qualitatively, the \textit{H}-like aggregate exhibits broad positive and negative potential lobes with a relatively large spatial separation (Fig.~\ref{fig5} (a), right), whereas the \textit{I}- and \textit{J}-like aggregates gradually develop more localized dipolar potential with reduced spatial extent (Fig.~\ref{fig5} (b) and (c), right).

To quantify the observed electrostatic localization, we analyze the screened transition electrostatic potentials using a series of common absolute potential thresholds. Across the entire range of thresholds, the \textit{H}-like packing consistently exhibits the most spatially extended screened electrostatic potential, while the \textit{J}-like cluster remains the most compact. At the visualization threshold of $|\Phi_{\text{tr}}^{\text{scr}}|=10$ a.u. used in Fig.~\ref{fig5}, the centroid separation between the positive and negative potential decreases from 54.6 $\text{\AA}$ for the \textit{H}-like aggregate cluster to 41.3 $\text{\AA}$ for the \textit{J}-like aggregate cluster, corresponding to an approximately $24\%$ reduction. Simultaneously, the volume of the high-magnitude potential region decreases from 4,671 to 1,660 $\text{\AA}^3$, representing nearly a threefold contraction (see the SI for detailed analysis). The packing-dependent variation of $\Phi_\text{tr}^\text{scr}(\boldsymbol{r})$ provides a real-space representation of how dielectric screening redistributes the transition-induced electrostatic potential. Its direct photophysical consequences are assessed below through the screened Frenkel couplings and excitonic band structure analysis.

\subsection{Real-space validation of the Frenkel exciton picture}
A central assumption of the Frenkel exciton model is that an aggregate excitation can be represented as a coherent superposition of localized monomer excitations.\cite{Czader2008,Krieger2025,Davydov1964,Spano2010,Hestand2018,Deshmukh2019,Chuang2019,Deshmukh2020,Deshmukh2022} Since sBSE calculations provide the transition densities of both the isolated monomer and the aggregate clusters, we directly test this assumption in real space by reconstructing the aggregate transition density from rigidly translated copies of the monomer ones.\cite{Czader2008,Krieger2025}

\begin{table}[h]
\caption{Real-space validation of the Frenkel representation in $4\times3$ \textit{H}-, \textit{I}-, and \textit{J}-like aggregate clusters. Definitions of the projected overlap, and equal-weight overlap are provided in the Computational Methods section and in the SI.}
\label{tab:frenkel_validation}
\centering
\begin{tabular}{ccc}
\toprule
Motif & Projected Overlap & Equal-weight Overlap \\
\midrule
\textit{H} & 94.3\% & 92.0\% \\
\textit{I} & 93.8\% & 89.7\% \\
\textit{J} & 92.5\% & 86.4\% \\
\bottomrule
\end{tabular}
\end{table}

As summarized in Table~\ref{tab:frenkel_validation} and detailed in the SI, the projected overlap exceeds $92.5\%$ for all three $4\times3$ motifs, demonstrating that the reconstructed transition densities closely reproduce the many-body excitation. By comparison, the equal-weight overlap ranges from $86.4-92.0\%$, showing that optimizing the Frenkel coefficients provides a slight improvement over a uniform distribution. Meanwhile, the condition number in the projected overlap process remains close to unity, confirming that the mapped monomer transition densities form a stable projection basis. The remaining $10-15\%$ of the transition density lies outside the Frenkel basis and likely reflects local polarization, subtle transition-density deformation, and effects beyond the one-excitation-per-monomer picture.

\subsection{Spatially dependent screening of Frenkel excitons}
\begin{figure}[H]
\centering
\includegraphics[width=1\textwidth]{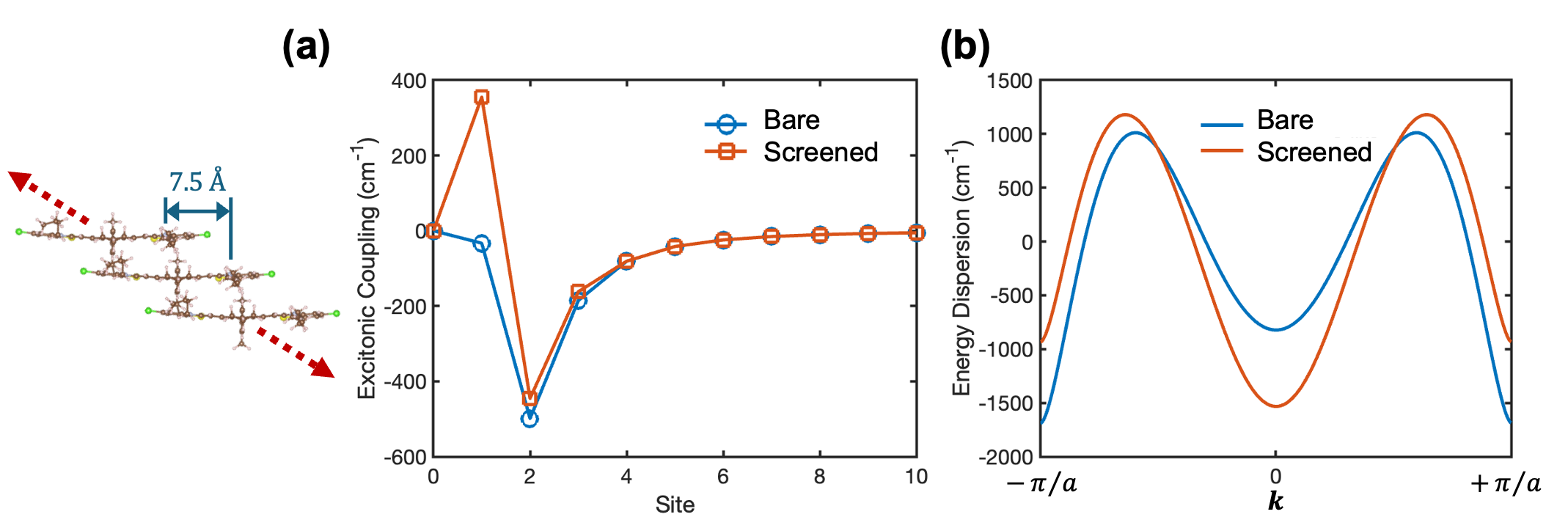}
\caption{\textbf{(a)} Coupling $J_{nm}$ as a function of inter-site distance $|\boldsymbol{r}_n-\boldsymbol{r}_m|$ in a 1D chain with a 7.5 $\text{\AA}$ slip. \textbf{(b)} The excitonic band structures calculated from the unscreened and screened (derived from the $J$-like $4\times1$ cluster) couplings for the same 1D aggregate.}
\label{fig:frenkel1D}
\end{figure}

Having established that the aggregate sBSE transition densities are well represented within the Frenkel basis, we further assess the consequences of the screened couplings in the context of excitonic molecular aggregates. Fig.~\ref{fig:frenkel1D} (a) compares $J_n$, the excitonic coupling between sites separated by $n$ lattice spacings, in a 1D chain before and after screening for a given lattice where the screening effect is most dramatic. As expected, screening has negligible effects on further-separated sites $n>3$, showing the near-field nature of such screening. On the other hand, screening becomes more significant between near neighbors. In fact, for the current example, the unscreened nearest-neighbor coupling $J_1$ is vanishing while the screened one is significantly nonzero. This is far from the expectation of a spatially homogeneous screening constant picture. The packing dependence can be understood in terms of geometry-dependent cancellation between positive and negative transition-density contributions, which can strongly modify near-field excitonic couplings and their distance scaling.\cite{Chuang2014}

This unusual observation of ``anti-screening" results from the combination of the spatially extended transition-charge distribution and range-sensitive screening. In the bare coupling, contributions from different transition-charge pairs largely cancel one another, whereas short-range screening acts differently on the individual contributions and disrupts this cancellation, leaving a strong nonzero cumulative coupling. A detailed analysis of this effect and its dependence on molecular slip is provided in the SI.

We also investigate the screening's effect on the excitonic band structure in Fig.~\ref{fig:frenkel1D} (b). At this particular slip value (7.5 $\text{\AA}$), the unscreened interactions lead to a band-edge \textit{J}-aggregate where the bright state ($\boldsymbol{k}=0$ in the dipole, long-wavelength regime) is red-shifted compared to monomers and located at the bottom of the exciton band.\cite{Chuang2019,Deshmukh2022} This is in contrast to the screened case, where the bright state is blue-shifted above the band edge, and the lowest-energy exciton state is of high quasimomentum ($|\boldsymbol{k}|=\pi/a$, where $a\sim25$ $\text{\AA}$ is the 1D lattice constant for Cy7-DiMePh) and dark in nature. This is a direct result of significantly changing the nearest-neighbor coupling from vanishing to strongly positive, thereby qualitatively modifying the excitonic band structure. Consequently, spatially dependent screening can alter not only the magnitude of excitonic coupling but also the energetic ordering and optical character of the aggregate exciton states.

We note that similar effects apply to 2D aggregates as well, albeit with more complication, as a variety of selective screening schemes similar to that described in the above 1D case may act in concert with or against one another depending on the specific packing condition. It is, however, expected that accounting for such screening effects is crucial in predicting accurate excitonic coupling and the overall photophysics of these aggregates.

\section{Computational Methods}\label{sec3}

\subsection{Ground-state orbitals and near-gap hybrid reference}
We first perform grid-based, norm-conserving pseudopotential, local-density approximation (LDA) density functional theory calculations to obtain an initial ground-state reference.\cite{Perdew1992,Troullier1991} A near-gap hybrid (ngH) calculation with the CAM-LDA0 range-separated hybrid functional ($\alpha=0.19,\beta=0.46,\gamma=0.33$ $\text{Bohr}^{-1}$) is then used to improve the frontier-orbital energetics with explicit hybrid exchange added in.\cite{Yanai2004,Bradbury2023a} In the ngH framework, only a subset containing $N_v$ occupied and $N_c$ unoccupied LDA
molecular orbitals (MOs) is used, with $N_c=3N_v$ as a typical choice. See the SI for more information. Within this orbital basis, non-local exchange is evaluated using a mixed deterministic-stochastic approach. The range-separated Coulomb interaction is partitioned into low- and high-momentum components, with the small low-momentum space treated deterministically and the much larger high-momentum space compressed into a sparse stochastic basis.

\subsection{Stochastic sampling of the screened interaction}

The excited-state calculations are carried out using sBSE, previously denoted as TDHF@$v_W$.\cite{Bradbury2022,Bradbury2024,Li2025} In the static limit, the screened Coulomb interaction entering the electron-hole kernel is the zero-frequency screened interaction,

\begin{equation}
    W(\boldsymbol{r},\boldsymbol{r}';\omega=0) = \int\epsilon^{-1}(\boldsymbol{r},\boldsymbol{r}'';\omega=0)v(\boldsymbol{r}'',\boldsymbol{r}')\,d\boldsymbol{r}'',
\label{eq:W_full_real}
\end{equation}
where $\epsilon^{-1}(\boldsymbol{r},\boldsymbol{r}'';\omega=0)$ is the static inverse dielectric matrix and $v(\boldsymbol{r}',\boldsymbol{r}'')=|\boldsymbol{r}'-\boldsymbol{r}''|^{-1}$ is the bare Coulomb potential. Direct construction of $W=\epsilon^{-1}v$ requires an explicit dielectric response matrix, which is prohibitive for large aggregates. Instead, we estimate the action of the screened interaction using stochastic occupied-pair densities. Several established approaches avoid the explicit construction and inversion of the dielectric matrix by evaluating the screened interaction or dielectric response through alternative iterative, linear-response, or time-propagation formulations.\cite{Allen2026b,Rabani2015,Ren2012,Rocca2012,Wilson2008}

A stochastic occupied-pair density, $\beta(\boldsymbol{r})$, is constructed from products of independent stochastic occupied orbitals such that statistical averaging recovers the full occupied-pair space. For each $\beta$, the electrostatic potential produced from the random charge distribution  acts as an impulsive perturbation on the Fermi sea of electrons. The induced density response is then obtained from short-time linear-response time-dependent Hartree propagation.\cite{Bradbury2023b} Rather than explicitly constructing the full two-point $W(\boldsymbol{r},\boldsymbol{r}';\omega=0)$, we evaluate its action on each stochastic density, $W_\beta(\boldsymbol{r})$, and use this ensemble to construct a translationally invariant effective screening kernel, $v_W(\boldsymbol{r}-\boldsymbol{r}')$, or equivalently $v_W(\boldsymbol{k})$ in reciprocal space. Full definitions and derivations are provided in Refs. \cite{Bradbury2022,Bradbury2023b,Li2025,Allen2026} and the SI.

\subsection{Polynomial-Gaussian parameterization of \texorpdfstring{$v_W$}{vW}}
First, we define $g(\boldsymbol{k})\equiv\epsilon^{-1}(\boldsymbol{k})-1$, so that $\epsilon^{-1}(\boldsymbol{k})=1+g(\boldsymbol{k})$ and $v_W(\boldsymbol{k})=v(\boldsymbol{k})g(\boldsymbol{k})$. Here, $v(\boldsymbol{k})=4\pi/k^2$ is the bare Coulomb interaction. We expand $g(\boldsymbol{k})$ in a finite polynomial-Gaussian basis,

\begin{equation}
    g^{\text{param.}}(\boldsymbol{k})=\sum_{m=1}^{N_{\text{basis}}}
    \theta_m f_m(\boldsymbol{k}),
\label{eq:g_expand_main}
\end{equation}
where $\{\theta_m\}$ are parameterization coefficients and $\{f_m(\boldsymbol{k})\}$ are predefined basis functions. The basis contains isotropic Gaussian terms together with anisotropic polynomial-Gaussian functions that capture directional screening along the molecular and stacking directions. See the SI for more details.

For each basis function, $f_m(\boldsymbol{k})$, we define the corresponding interaction kernel,

\begin{equation}
    v_m(\boldsymbol{k})=\frac{4\pi}{k^2}f_m(\boldsymbol{k}).
\label{eq:vm}
\end{equation}
The action of this kernel on a stochastic occupied-pair density,
$\beta_n(\boldsymbol{r})$, is $(v_m\beta_n)(\boldsymbol{r})$ i.e., a convolution of $v_m(\boldsymbol{r}-\boldsymbol{r}')$ with $\beta_n(\boldsymbol{r}')$, or equivalently, $v_m(\boldsymbol{k})\beta_n(\boldsymbol{k})$ in reciprocal space. For $\beta_n(\boldsymbol{r})$, the parameterized screened response is written as

\begin{equation}
    v_W^{\text{param}}\beta_n
    =\sum_{m=1}^{N_{\text{basis}}}\theta_mv_m\beta_n.
\label{eq:model_response_main}
\end{equation}
The coefficients are obtained from a least-squares minimization over all stochastic samples, i.e., 

\begin{equation}
    \mathcal{J}(\boldsymbol{\theta})=\sum_n\left\|W\beta_n-\sum_m\theta_mv_m\beta_n\right\|^2,
\label{eq:loss_main}
\end{equation}
which reduces to a simple linear system, $\sum_n A_{mn}\theta_n=b_m$.

It is possible that the finite polynomial-Gaussian expansion introduces
small fitting artifacts, leading to local deviations of the parameterized
inverse dielectric response outside the target range of $0<\epsilon^{-1}(\boldsymbol{k})\le1$. These excursions are smoothly
suppressed using an error-function filter, as detailed in the SI. The final parameterized $\epsilon^{-1}(\boldsymbol{k})$ is easily converted to $v_W(\boldsymbol{k})$ and used in all production sBSE calculations reported in this work. The polynomial-Gaussian parameterized $v_W$ captures $\sim82\%-88\%$ of the static $W$, similar to the previously reported all-$\boldsymbol{k}$ fit. However, this parameterized kernel is transferable and grid-independent, with only 16 parameters.\cite{Bradbury2023b,Li2025,Allen2026}

\subsection{sBSE spectra and excited-state information}

Excited-state calculations are performed using the full resonant-antiresonant sBSE Liouvillian, with screened exchange represented by the polynomial-Gaussian $v_W$ kernel described above. Optical spectra are evaluated without explicit diagonalization using a Chebyshev expansion of the spectral projector, with polarization-resolved dipole excitation vectors constructed along the $x$, $y$, and $z$ directions.\cite{Weisse2006,Bradbury2022,Bradbury2023b,Sereda2024} Detailed expressions for the Liouvillian superoperator, dipole vectors, absorption cross section, and Chebyshev recursion are provided in the SI.

Energy-filtered exciton amplitudes extracted from the Chebyshev representation are used to characterize the composition and spatial extent of the excited states. Exciton delocalization is quantified through the participation ratio, while the corresponding real-space transition density, $n_{\text{tr}}(\boldsymbol{r};\omega)$, is reconstructed from the filtered occupied-virtual amplitudes.\cite{Allen2026} This transition density is then used to construct both the bare transition electrostatic potential and its screened counterpart using the same nonlocal dielectric response employed in the sBSE Hamiltonian.\cite{Allen2026} Full definitions and equations for the participation ratio, transition density, and bare and screened transition electrostatic potentials are given in the SI.

\subsection{Real-space mapping of the Frenkel exciton picture}

To directly assess the Frenkel exciton picture in real space, the bright-state transition density defined above is calculated for both the isolated monomer and the aggregate. The monomer transition density, $n_{\text{tr}}^{\text{mono}}(\boldsymbol{r})$, is rigidly mapped onto each molecular site of the aggregate using the corresponding rotation, $R_i$, and translation, $\boldsymbol{t}_i$,

\begin{equation}
    n_\text{tr}^{(i)}(\boldsymbol{r})=
    n_{\text{tr}}^{\text{mono}}
    \left[R_i^{-1}(\boldsymbol{r}-\boldsymbol{t}_i)\right].
\end{equation}

The aggregate sBSE transition density is then projected onto this localized monomer transition-density basis to obtain the Frenkel coefficients and the corresponding reconstructed transition density. The quality of the Frenkel representation for the bright state considered here is quantified by its normalized overlap with the full aggregate transition density.\cite{Czader2008,Krieger2025} For comparison, we also evaluate an equal-weight Frenkel reconstruction in which all monomers contribute with identical amplitude. Detailed expressions for the least-squares projection, normalization, projected overlap, and equal-weight overlap are provided in the SI.

\subsection{1D Frenkel exciton model with screened couplings}

To further assess the consequences of spatially dependent screening, we consider a 1D molecular aggregate within the one-exciton Frenkel manifold, where the basis state $|n\rangle$ represents the $n$-th site in the excited state while all other sites remain in the ground state. The exciton Hamiltonian is

\begin{equation}
    H_\text{ex}=\sum_n \varepsilon_n |n\rangle\langle n|+\sum_{n\neq m}J_{nm}|n\rangle\langle m|,
\end{equation}
where $\varepsilon_n$ is the transition energy of the $n$-th site and $J_{nm}$ is the excitonic coupling between the sites $n$ and $m$. To account for near-field effects, the coupling is evaluated using a transition-monopole representation,

\begin{equation}
    J_{nm}=\sum_{k=1}^{N_m}\sum_{l=1}^{N_n}\epsilon^{-1}(\boldsymbol{r}_{kl})\frac{q_k^m q_l^n}{r_{kl}},
\end{equation}
where $\epsilon^{-1}(\boldsymbol{r}_{kl})$ is the real-space inverse dielectric screening factor, $q_k^m$ is the transition charge on atom $k$ of molecule $m$, and $\boldsymbol{r}_{kl}$ and $r_{kl}=|\boldsymbol{r}_{kl}|$ denote the separation vector and distance, respectively, between atoms $k$ and $l$.

\section{Conclusion}\label{sec4}

We have applied an \textit{ab initio} sBSE approach to Cy7-DiMePh molecular aggregates containing up to 2,856 valence electrons. By replacing the explicit screened Coulomb interaction with a transferable, polynomial-Gaussian parameterized screened kernel, we obtained BSE-level excited-state information for aggregate sizes that are beyond the reach of conventional BSE calculations.

The central result is that aggregation changes not only the exciton coupling but also the dielectric screening itself. The optical spectra follow the expected aggregate ordering, with \textit{H}-like packing producing a small blueshift, \textit{I}-like cluster showing a gradual redshift, and \textit{J}-like packing producing a large redshift into the SWIR range. PRs and transition densities show that aggregation converts the monomer excitation into collective excitons distributed over many occupied-virtual transitions and multiple chromophores. Beyond the transition densities, the corresponding screened transition electrostatic potentials provide an intuitive real-space picture of the excitonic screening. As the packing evolves from \textit{H}- to \textit{J}-like configurations, the screened transition potential becomes progressively more compact, exhibiting reduced high-potential volume and smaller positive-negative lobe separation. This electrostatic localization accompanies the exciton redshift and establishes a direct connection between molecular packing, dielectric screening, and real-space organization. Unlike transition-dipole models,\cite{Czader2008,Krieger2025,Davydov1964,Spano2010,Hestand2018,Deshmukh2019,Chuang2019,Deshmukh2020,Deshmukh2022,Williams2024} the present approach makes no assumptions regarding site energies or pairwise couplings. Instead, both the screened electron-hole interaction and the resulting transition electrostatic potentials emerge directly from the many-body electronic response of the aggregate. Remarkably, the resulting sBSE transition density is accurately represented by a coherent superposition of localized monomer transition densities, demonstrating that the Frenkel picture naturally emerges from the many-body solution. Within this representation, spatially dependent screening selectively modifies short-range excitonic couplings and can qualitatively alter the resulting band structure and bright-state ordering.

Several directions follow naturally from this work. First, the same framework can be applied to tubular cyanine aggregates, where curvature and helical packing produce screening patterns not present in planar systems.\cite{Megow2015,Williams2024,Bailey2023} Second, aggregate disorder and solvent environments can be incorporated to connect more directly with experimental spectra.\cite{Tempelaar2014,Hestand2018,Chuang2019,Deshmukh2020,Bradbury2020} Finally, extending the method to periodic boundary conditions would enable direct treatment of extended molecular crystals.

\vspace{0.5in}

\backmatter

\bmhead{Supplementary Information}
A Supplementary Information (SI) file is available for this article. The SI includes monomer validation, aggregate construction procedures, computational details, derivation of the parameterized screened interaction, details of the sBSE formalism and spectral evaluation, validation of the parameterized $v_W$ representation against the form-free screened interaction, $v_W$ parameters for representative systems, screened transition electrostatic potential analysis, Frenkel exciton analysis, and additional spectra illustrating aggregate-size evolution.

\bmhead{Acknowledgements}
We acknowledge the computational resources provided by the Expanse cluster at the San Diego Supercomputer Center (SDSC) through allocation CHE-260063 under the Advanced Cyberinfrastructure Coordination Ecosystem: Services \& Support (ACCESS) program. BYL is grateful for support from the Azrieli Foundation through the Azrieli International Postdoctoral Fellowship. CC acknowledges support from the University of Nevada, Las Vegas (UNLV). TD is supported by the NSF Graduate Research Fellowship Program. JRC appreciates support from NSF Grant CHE-2506132, and DN is supported by NSF Grant CHE-2245253 and BSF Grant BSF-2024265.

\bmhead{Author Contributions}
BYL and CC contributed equally to this work. BYL developed the parameterized screened-interaction framework, constructed the aggregate geometries, and performed the sBSE calculations. CC developed and performed the Frenkel exciton analysis, including the transition-density projection, screened excitonic coupling calculations, and excitonic band-structure analysis. BYL and CC analyzed the data, generated the figures, and wrote the original manuscript. TD and TA performed data verification and convergence validation and contributed to manuscript editing. JRC contributed to project supervision, interpretation of the results, and manuscript editing. DN, TA, and BYL developed the underlying stochastic framework and computational codes; DN supervised the project and contributed to manuscript editing. All authors discussed the results and approved the final manuscript.

\bmhead{Competing Interests}
The authors declare no competing interests.

\bmhead{Correspondence and Requests for Materials}
Should be addressed to Barry Y. Li and Chern Chuang.

\vspace{0.5in}

\bibliography{sn-bibliography}

\clearpage


\section*{Supplementary material}

# Contents

# 1 Geometries

## 1.1 Cy7-DiMePh monomer geometry and electronic structure

The Cy7-DiMePh monomer is optimized using the ORCA 6.0 package at the B3LYP/def2-TZVPPD level with the D4 dispersion correction and an implicit methanol environment.[1–5] The resulting full Cartesian coordinates are listed below. The terminal sulfonate groups of the experimental dye are replaced by methyl groups for simplicity. This choice preserves the electronic structure of the polymethine chromophore. The optimized structure exhibits the characteristic cyanine architecture consisting of a conjugated heptamethine bridge connecting two heterocyclic end groups and a central 3,5-dimethylphenyl substituent.[6, 7]

Fig. S1 compares the highest occupied molecular orbitals (HOMOs) and lowest unoccupied molecular orbitals (LUMOs) obtained from grid-based local-density approximation (LDA) and the near-gap hybrid (ngH) CAM-LDA0 Hamiltonian used throughout the sBSE calculations. In both cases, the HOMO and LUMO remain strongly delocalized along the polymethine backbone, indicating that the low-energy optical transitions are dominated by the extended $\pi$-conjugated framework.

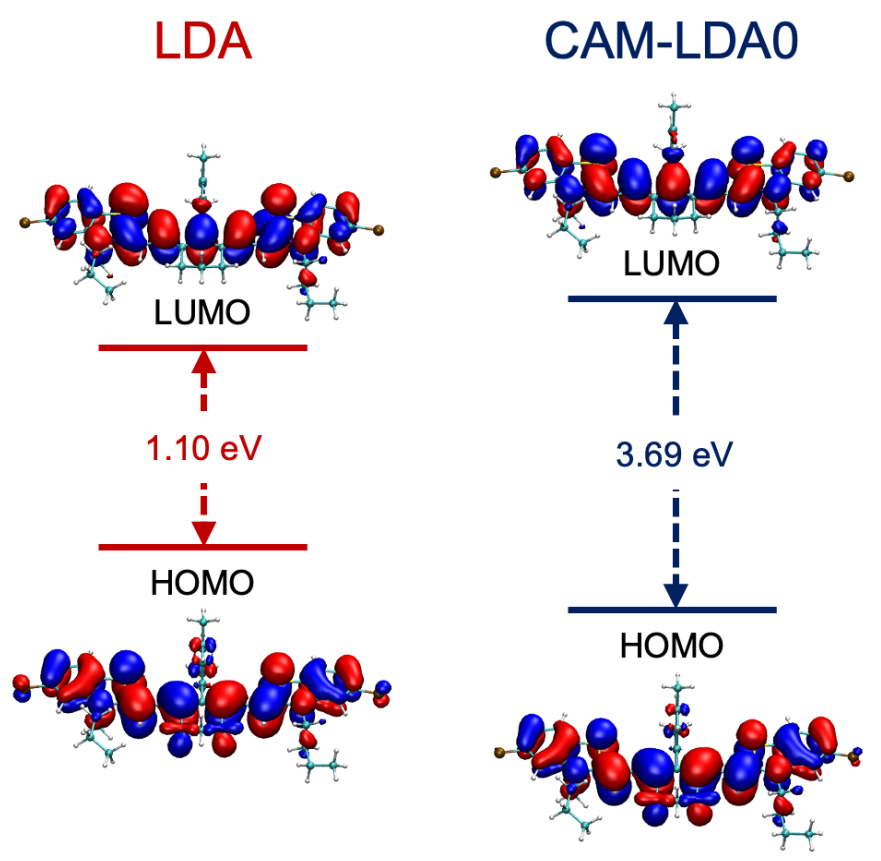


**Fig. S1** HOMOs and LUMOs of the Cy7-DiMePh monomer calculated using grid-based LDA and ngH CAM-LDA0. HOMO-LUMO gaps are indicated in eV.

89

Charge = +1, 2S+1 = 1

C -4.80728185959975 0.46082564220115 -0.46772717392124
C -3.70840866375433 -0.36051897104414 -0.24607655013954
C -2.42012255993716 0.13648605033317 -0.09158759277093
C -1.26520115341928 -0.61416521577321 0.12103651738403
C -0.01444925006394 0.03123783247986 0.19140907001600
C 1.19202143747918 -0.68576967375387 0.31844999297169
C 2.40797115235338 -0.00624279241220 0.27297871120446
C 3.67369806027698 -0.57764759165306 0.31832729788738
C 4.83781845803628 0.17453260844845 0.21561731406180
C -1.33747265510561 -2.11730503558827 0.23661051741499
C -0.16246504345716 -2.67986817419507 1.02573705044325
H -1.35279281447728 -2.56092250514079 -0.76591179191062
H -2.27399750535267 -2.40930024504823 0.71323358576695
C 1.16002654077568 -2.18894231466377 0.45214258180984
H 1.98290825932811 -2.52227365489192 1.08603078140822
H 1.32799676528300 -2.64466523284495 -0.53067550012753
H -0.19299015498242 -3.77064763761032 1.01977752972267
H -0.24405669496418 -2.36233321231579 2.06946269055403
H -3.88603325808707 -1.42313033860627 -0.19420022396519
H -2.29035457097862 1.20970025611124 -0.15659086338633
H 2.35240206548571 1.07053318543034 0.17140756675677
H 3.78045341642576 -1.64692101752825 0.41132772407835
N -6.08453210225632 0.03421900938819 -0.64377950769699
S -4.71602905379388 2.20706645695317 -0.55061082124667
C -6.44055486775476 2.32166574555322 -0.79311042935225
C -7.01837019630228 1.04776848423675 -0.80714684646302
S 4.85310775913875 1.90889388318491 -0.02470640554886
C 6.59712571488409 1.92216888282921 -0.08136752600990
C 7.10175633902227 0.62809854982213 0.08365495625429
N 6.10244728056197 -0.31734829445091 0.27023963641027
C -7.20798772725622 3.46332332271830 -0.95373441312774
C -8.57864031165134 3.32522874293866 -1.12676695815519
C -9.14128795843206 2.05201061986540 -1.13253246665064
C -8.39092275530576 0.89765517497672 -0.97395609197586

C 7.43877248388427 3.00653727091676 -0.26552780759624
C 8.80964395371908 2.78800638196856 -0.29292108380711
C 9.29796478161648 1.49368350061165 -0.13859704294711
C 8.47284184314394 0.39620719504820 0.04912854623245
H -6.75728145583701 4.44594753761344 -0.94410960868353
H -9.20132559330040 4.19862793139270 -1.25261223538723
H -8.86193934390180 -0.07264396480296 -0.97738403648985
H 8.88644714433339 -0.59398104603571 0.15645349232946
H 7.04458692202433 4.00566351067811 -0.38842981292643
H 9.48925829748808 3.61496264597024 -0.43685506715734
Cl -10.87115593226966 1.89729915555444 -1.34482530871031
Cl 11.02761994780082 1.23485889694570 -0.18859422986061
C 0.03656227669881 1.51957095986949 0.09597994550182
C 0.12236871952047 2.14309839801720 -1.14984723421759
C 0.17166952277762 3.52950264065108 -1.25161528492229
C 0.13286992890058 4.28728804640827 -0.07752794700768
C 0.04663552107245 3.68935757289419 1.17568387213679
C -0.00075698506539 2.29457532382861 1.24804715038281
H 0.17066419130762 5.36886332519866 -0.14583607856734
H 0.15118256163785 1.53664173358963 -2.04675225886761
H -0.06759212251234 1.80971835686268 2.21478503322156
C -6.45875736043259 -1.38286739129817 -0.59136667180163
C -6.73171839329123 -1.84751714474109 0.83639100207938
H -5.65964322146225 -1.95026117583112 -1.05991549268318
H -7.34053145685759 -1.51100630066135 -1.21502747007851
C -7.07620748795327 -3.33561723488253 0.91628554312010
H -7.55566647437482 -1.25527547000223 1.24089641350274
H -5.85465327403164 -1.63811689882530 1.45464216904696
C 6.39705194706212 -1.74302102328901 0.45183611564736
C 6.43439568644155 -2.49358733381537 -0.87644673228570
H 5.64345022434731 -2.15979332177362 1.11669966539433
H 7.34586035674289 -1.80719059633754 0.97743257789342
C 6.69183220515060 -3.99074957407641 -0.70039980128666
H 7.20814749845180 -2.05483386987622 -1.51239185807216
H 5.48033861601305 -2.34495207130620 -1.38715415043395
C 0.26276484886009 4.20611440948891 -2.59172151658453

C 0.00621163188003 4.51494382516779 2.43223903773819

C -5.91579691810879 -4.26264191503606 0.55857888074982

H -7.93079399358454 -3.54546225721270 0.26536790059665

H -7.40703410092135 -3.55359542889946 1.93475183768279

H -5.60510091642769 -4.15083466910199 -0.48183419195799

H -6.19549954768606 -5.30828167832688 0.70263941156238

H -5.04535200378314 -4.06262706551462 1.18873161710746

C 8.07560823185203 -4.32782461334039 -0.14868745097970

H 6.56756600502887 -4.47320336322289 -1.67313680015951

H 5.92127256847330 -4.41383482861636 -0.04818955443931

H 8.86253683390529 -3.90754963618954 -0.78033029624626

H 8.22526961473461 -5.40870928204437 -0.10871147227783

H 8.22025532950043 -3.94224369313536 0.86243251787764

H 0.29961320937577 3.47819636221360 -3.40208752612934

H 1.15584977281854 4.83203043511360 -2.65867609269891

H -0.59698993268481 4.85843151544811 -2.76362438902762

H 0.05665653229722 5.58054863803861 2.20904590443838

H 0.84065302786171 4.26810915550027 3.09309026866330

H -0.91152781435571 4.33097558325504 2.99612523968880

## 1.2 Aggregate construction and geometry protocol

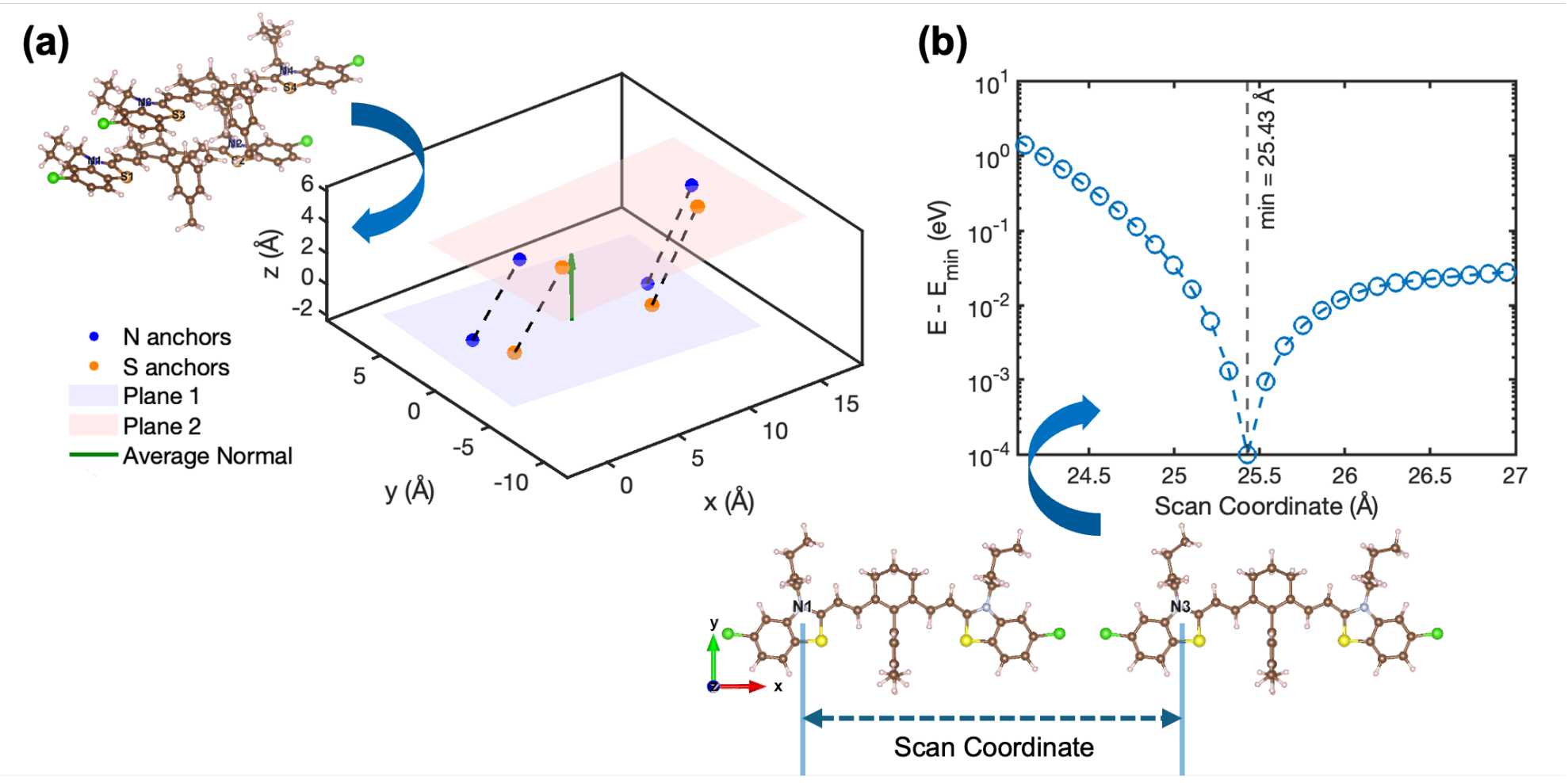


**Fig. S2** **(a)** Definition of intermolecular $\pi$-$\pi$ stacking geometry used to construct aggregate configurations. Molecular planes are defined using sulfur and nitrogen anchor atoms, and average plane normals are used to characterize the inter-layer distance of 4.72 Å. **(b)** Potential energy surface scan to find the lateral spacing between adjacent molecules. The scanning coordinates are the N−N distance between two molecules along their "long axis". A logarithmic scale is used for the total energy difference, $E - E_{\mathrm{min}}$. An optimal distance of 25.43 Å is identified.

Aggregate structures are constructed from the optimized monomer geometry. As illustrated in Fig. S2 (a), the intermolecular $\pi$-$\pi$ stacking distance is determined from a constrained dimer optimization. Molecular planes are defined using sulfur and nitrogen anchor atoms, and average plane normals are used to characterize the relative orientation of neighboring chromophores. The optimized dimer yields an equilibrium inter-layer separation of approximately 4.72 Å, which was subsequently adopted for all aggregate constructions.

The lateral displacement along the molecular long (N−N) axis is determined by a potential energy surface scan by translating one monomer relative to the other. As shown in Fig. S2 (b), the total energy exhibits a clear minimum at 25.43 Å. Therefore, this distance was used as the equilibrium lateral spacing for aggregate construction. Both the constrained dimer optimization and the lateral potential energy surface scan

are performed in the ORCA 6.0 package using the $r^2$SCAN-3c method with an implicit water solvent.[4, 5, 8]

Using the optimized $\pi$-$\pi$ separation and lateral spacing, different aggregate clusters are generated by systematically varying the intermolecular slip along the molecular "long axis." Slip distances of 2.60 Å, 4.60 Å, and 6.60 Å are employed to represent $H$-, $I$-, and $J$-like packing motifs, respectively. These values span the experimentally relevant range reported for cyanine aggregates and provide a controlled structural series.[9–12]

## 2 General Information for Computation

The computational parameters used for all systems are summarized in Table S1.

**Table S1** Grid and active-space parameters used in this work.

| | **Monomer** | **Dimer 2 × 1** | | | **Trimer 3 × 1** | | | **Tetramer 4 × 1** | | | **Aggregate 4 × 3** | | |
|---|---|---|---|---|---|---|---|---|---|---|---|---|---|
| | | $H$ | $I$ | $J$ | $H$ | $I$ | $J$ | $H$ | $I$ | $J$ | $H$ | $I$ | $J$ |
| $dx$ | 0.5 | 0.5 | 0.5 | 0.5 | 0.5 | 0.5 | 0.5 | 0.5 | 0.5 | 0.5 | 0.5 | 0.5 | 0.5 |
| $n_x$ | 108 | 118 | 126 | 132 | 128 | 142 | 158 | 138 | 160 | 182 | 326 | 350 | 372 |
| $n_y$ | 66 | 66 | 66 | 66 | 66 | 66 | 66 | 66 | 66 | 66 | 66 | 66 | 66 |
| $n_z$ | 50 | 68 | 68 | 68 | 86 | 86 | 86 | 102 | 102 | 102 | 102 | 102 | 102 |
| $N_v$ | 40 | 80 | 80 | 80 | 120 | 120 | 120 | 160 | 160 | 160 | 240 | 240 | 240 |
| $N_c$ | 120 | 240 | 240 | 240 | 360 | 360 | 360 | 480 | 480 | 480 | 720 | 720 | 720 |
| $N_{k\text{-low}}$ | 749 | 1123 | 1225 | 1249 | 1515 | 1721 | 1927 | 1959 | 2271 | 2599 | 4649 | 4969 | 5305 |

Real-space computational grids, where a uniform grid in Bohr is used, i.e., $dx = dy = dz$. $N_v$ and $N_c$ denote the number of occupied and virtual orbitals included in the near-gap active space, respectively. $N_{k\text{-low}}$ is the number of deterministically treated long-wavelength terms, and the high-$k$ space is represented with a similar number of sparse stochastic vectors. Details in Ref. [13].

# 3 Details of the Parameterized Screened Interaction and sBSE Formalism

## 3.1 Stochastic occupied-pair sampling

A stochastic valence occupied orbital is constructed as

$$\bar{\beta}(\boldsymbol{r}) = \sum_{i \in N_v} \bar{\beta}_i \phi_i(\boldsymbol{r}), \qquad \bar{\beta}_i = \pm 1, \tag{S1}$$

and another statistically independent orbital is,

$$\bar{\bar{\beta}}(\boldsymbol{r}) = \sum_{j \in N_v} \bar{\bar{\beta}}_j \phi_j(\boldsymbol{r}), \qquad \bar{\bar{\beta}}_j = \pm 1, \tag{S2}$$

Together these orbitals form the occupied-pair density

$$\beta(\boldsymbol{r}) = \bar{\beta}(\boldsymbol{r})\bar{\bar{\beta}}(\boldsymbol{r}). \tag{S3}$$

Statistical averaging over a large number of stochastic realizations recovers the deterministic occupied-pair density,

$$\{\beta(\boldsymbol{r})\} = \left\{\bar{\beta}_i \bar{\bar{\beta}}_j \bar{\beta}(\boldsymbol{r})\bar{\bar{\beta}}(\boldsymbol{r})\right\} = \phi_i(\boldsymbol{r})\phi_j(\boldsymbol{r}) \tag{S4}$$

The above expression uses $\{\bar{\beta}_i \bar{\beta}_j\} = \{\bar{\bar{\beta}}_i \bar{\bar{\beta}}_j\} = \delta_{ij}$ in limit $N_\beta \to \infty$. In practice, a small number of samples is sufficient to converge the screened-exchange kernel.

For each stochastic density, the bare perturbing potential is

$$v_\beta(\boldsymbol{r}) = \int v(\boldsymbol{r} - \boldsymbol{r}')\beta(\boldsymbol{r}')\, d\boldsymbol{r}', \tag{S5}$$

and the induced density, $\Delta n_\beta(\boldsymbol{r})$, is obtained by short-time linear-response propagation.[14, 15] The corresponding screened-response potential is

$$W_\beta(\boldsymbol{r}) = \int v(\boldsymbol{r}-\boldsymbol{r}')\Delta n_\beta(\boldsymbol{r}')\,d\boldsymbol{r}'. \tag{S6}$$

## 3.2 Connection to the form-free screened interaction

In the original optimized attenuated interaction formalism,[7, 15, 16] the screened interaction is obtained by minimizing the functional

$$J = \sum_{ij} (\phi_i\phi_j|(W - v_W)^2|\phi_i\phi_j), \tag{S7}$$

where $W$ denotes the screened Coulomb interaction and $v_W$ is the effective screened exchange kernel. Replacing the orbital-pair products $\phi_i\phi_j$ by stochastic pair densities, $\beta(\boldsymbol{r})$, and moving to reciprocal space yields the form-free estimator,

$$v_W(\boldsymbol{k}) = \left\{ \frac{\beta^*(\boldsymbol{k})\langle \boldsymbol{k}|W|\beta\rangle}{|\langle \boldsymbol{k}|\beta\rangle|^2} \right\}_\beta, \tag{S8}$$

where the braces denote averaging over stochastic samples.

The form-free representation is mathematically attractive because each reciprocal-space grid point acts as an independent degree of freedom. In principle, this allows an exact representation of the screened interaction on the chosen reciprocal-space mesh. However, the resulting $v_W(\boldsymbol{k})$ is noisy in the large-$|\boldsymbol{k}|$ region and remains tied to a specific reciprocal-space grid. On the other hand, the previously employed one-dimensional seven-parameter parameterization combined with isotropic three-dimensional reconstruction is often too restrictive to reproduce the actual dielectric screening. Therefore, we develop here a compact isotropic+anisotropic representation that retains the physically relevant screening information while reducing stochastic noise and improving transferability.[17]

### 3.3 Polynomial-Gaussian parameterization

As shown in the main text, we expand $g(\boldsymbol{k})$ in a finite polynomial-Gaussian basis, where $N_{\text{basis}} = 16$. The basis functions are

$$\begin{aligned}
&f_1(\boldsymbol{k}) = e^{-k^2/0.1},\ f_2(\boldsymbol{k}) = e^{-k^2/0.3},\ f_3(\boldsymbol{k}) = e^{-k^2},\ f_4(\boldsymbol{k}) = e^{-k^2/3.0},\\
&f_5(\boldsymbol{k}) = k_x^2 e^{-k^2/0.3},\ f_6(\boldsymbol{k}) = k_y^2 e^{-k^2/0.3},\ f_7(\boldsymbol{k}) = k_z^2 e^{-k^2/0.3},\\
&f_8(\boldsymbol{k}) = k_x^2 e^{-k^2},\ f_9(\boldsymbol{k}) = k_y^2 e^{-k^2},\ f_{10}(\boldsymbol{k}) = k_z^2 e^{-k^2},\\
&f_{11}(\boldsymbol{k}) = k_x^4 e^{-k^2},\ f_{12}(\boldsymbol{k}) = k_y^4 e^{-k^2},\ f_{13}(\boldsymbol{k}) = k_z^4 e^{-k^2},\\
&f_{14}(\boldsymbol{k}) = k_x^2 k_y^2 e^{-k^2},\ f_{15}(\boldsymbol{k}) = k_x^2 k_z^2 e^{-k^2},\ f_{16}(\boldsymbol{k}) = k_y^2 k_z^2 e^{-k^2},
\end{aligned} \tag{S9}$$

with $k^2 = k_x^2 + k_y^2 + k_z^2$. The first four basis functions describe isotropic radial screening, and the remaining terms introduce anisotropy and mixed-direction couplings.

### 3.4 Difference from the form-free representation

In the form-free representation, every reciprocal-space grid point acts as an independent degree of freedom. For a reciprocal-space mesh of size $2n_x \times 2n_y \times 2n_z$ (the doubling comes from the use of the Martyna-Tuckerman approach for electron-electron interactions to reduce periodic images),[18] the screened interaction contains approximately $8\,n_x n_y n_z$ independent variables. The present parameterization replaces this by only $N_{\text{basis}} = 16$ coefficients, where $v_W(\boldsymbol{k}) \longrightarrow \{\theta_m\}_{m=1}^{16}$. More importantly, this parameterized kernel is transferable, i.e., it is no longer grid-dependent. With these 16 parameters, it can be reconstructed on any arbitrary grid for the same or different molecular systems.

### 3.5 Smooth physical filtering

The finite basis expansion may introduce small overfitting artifacts, particularly in regions where the stochastic signal is weak. As a result, the reconstructed inverse dielectric function may occasionally exceed the physically allowed interval of $0 < \epsilon^{-1}(\boldsymbol{k}) \leq 1$. To suppress such unphysical excursions, a smooth error-function filter is applied. First, the raw parameterization is constructed as

$$\epsilon^{-1}(\boldsymbol{k}) = 1 + g^{\text{param.}}(\boldsymbol{k}). \tag{S10}$$

The upper bound is enforced through

$$\epsilon_1^{-1}(\boldsymbol{k}) = \epsilon^{-1}(\boldsymbol{k}) - \frac{\epsilon^{-1}(\boldsymbol{k}) - 1}{2} \left[1 + \text{erf}\left(\alpha[\epsilon^{-1}(\boldsymbol{k}) - 1]\right)\right], \tag{S11}$$

which smoothly suppresses values exceeding unity. The lower bound is then imposed through

$$\epsilon_{\text{filt.}}^{-1}(\boldsymbol{k}) = \epsilon_1^{-1}(\boldsymbol{k}) - \frac{\epsilon_1^{-1}(\boldsymbol{k})}{2} \left[1 - \text{erf}\left(\alpha\, \epsilon_1^{-1}(\boldsymbol{k})\right)\right]. \tag{S12}$$

Throughout this work, $\alpha = 16$ is used. Finally,

$$g_{\text{filt.}}^{\text{param.}}(\boldsymbol{k}) = \epsilon_{\text{filt.}}^{-1}(\boldsymbol{k}) - 1. \tag{S13}$$

The filtering is applied only after the least-squares optimization and does not modify the fitting procedure itself.

### 3.6 sBSE spectra

The sBSE calculations are constructed from the ngH Hamiltonian,[1, 13] $H_{\text{ngH}}\phi_p(\boldsymbol{r}) = \varepsilon_p\phi_p(\boldsymbol{r})$, where $\phi_p(\boldsymbol{r})$ and $\varepsilon_p$ denote molecular orbitals and orbital energies, respectively. Occupied orbitals are labeled by $i, j$, virtual orbitals by $a, b$, and general orbitals

by $p, q$. Neutral excitations are represented in the occupied-virtual transition basis, $|ia\rangle \equiv \phi_i(\boldsymbol{r})\phi_a(\boldsymbol{r})$, with independent-particle excitation energies, $\Delta\varepsilon_{ia} = \varepsilon_a - \varepsilon_i$.

The sBSE excitation problem is solved in the occupied-virtual transition basis.[14, 15, 19, 20] We use the full resonant-antiresonant form rather than the Tamm-Dancoff approximation. The eigenvalue problem is

$$\mathcal{L}\begin{pmatrix} X \\ Y \end{pmatrix} = \begin{pmatrix} \mathcal{A} & \mathcal{B} \\ -\mathcal{B} & -\mathcal{A} \end{pmatrix}\begin{pmatrix} X \\ Y \end{pmatrix} = \Omega\begin{pmatrix} X \\ Y \end{pmatrix}, \tag{S14}$$

where $\mathcal{L}$ denotes the BSE Liouvillian, $\Omega$ is the excitation energy and $X$ and $Y$ are resonant and antiresonant amplitudes, respectively. The resonant block is

$$\mathcal{A}_{ia,jb} = \Delta\varepsilon_{ia}\delta_{ij}\delta_{ab} + K^{\text{dir.}}_{ia,jb} + K^{\text{sx.}}_{ia,jb}, \tag{S15}$$

and the coupling block is

$$\mathcal{B}_{ia,jb} = K^{\text{dir.}}_{ia,bj} + K^{\text{sx.}}_{ia,bj}. \tag{S16}$$

The direct Coulomb term for singlet excitations is

$$K^{\text{dir.}}_{ia,jb} = 2\iint \phi_i(\boldsymbol{r})\phi_a(\boldsymbol{r})v(\boldsymbol{r}-\boldsymbol{r}')\phi_j(\boldsymbol{r}')\phi_b(\boldsymbol{r}')d\boldsymbol{r}d\boldsymbol{r}'. \tag{S17}$$

The screened-exchange term is

$$K^{\text{sx.}}_{ia,jb} = -\iint \phi_i(\boldsymbol{r})\phi_j(\boldsymbol{r})v_W^{\text{param.}}(\boldsymbol{r}-\boldsymbol{r}')\phi_a(\boldsymbol{r}')\phi_b(\boldsymbol{r}')d\boldsymbol{r}d\boldsymbol{r}'. \tag{S18}$$

The parameterization allows the screened-exchange action to be evaluated without storing the full four-index interaction tensor.

Direct diagonalization of the sBSE Liouvillian is computationally expensive. Instead, we evaluate spectral quantities through a Chebyshev polynomial expansion of $\mathcal{L}$. For a given polarization direction, $\hat{r} = x, y, z$, we define the dipole excitation vector,

$$|\chi\rangle = \begin{pmatrix} \chi_{ia}^{+} \\ \chi_{ia}^{-} \end{pmatrix} = \begin{pmatrix} +\langle\phi_a|\hat{r}|\phi_i\rangle \\ -\langle\phi_a|\hat{r}|\phi_i\rangle \end{pmatrix}. \tag{S19}$$

The optical absorption spectrum is obtained from the spectral density associated with this dipole vector,

$$A(\omega) \propto \omega\langle\chi|\delta(\mathcal{L}-\omega)|\chi\rangle. \tag{S20}$$

To evaluate Eq. (S20), the Liouvillian is shifted and scaled to produce a dimensionless operator, $\widetilde{\mathcal{L}}$, whose spectrum lies within the interval $[-1, 1]$. The broadened spectral projector is then expanded in Chebyshev polynomials,[20–22]

$$\delta(\mathcal{L}-\omega) = \sum_{l=0}^{N_{\text{Cheb.}}} a_l(\omega) T_l(\widetilde{\mathcal{L}}), \tag{S21}$$

where $T_l$ is the $l^{\text{th}}$-order Chebyshev polynomial and $a_l(\omega)$ are the corresponding expansion coefficients. The absorption cross section is calculated from the Chebyshev moments,

$$\sigma(\omega) = \frac{4\pi\omega}{c} \sum_{l=0}^{N_{\text{Cheb.}}} a_l(\omega)\langle\tilde{\chi}|T_l(\tilde{\mathcal{L}})|\chi\rangle, \tag{S22}$$

where $c \simeq 137$ in atomic units, $|\tilde{\chi}\rangle = \pm|\chi\rangle$, and a simple decaying weight is used for the coefficients, $a_l(\omega)$. In our simulations, $N_{\text{Cheb.}} = 2,000$ is used across all systems. Thus, all spectra are evaluated with the same numerical broadening determined by the Chebyshev expansion; no explicit physical dephasing model is introduced. At a selected energy, $\omega$, the Chebyshev-filtered exciton amplitude is represented as $F_{ia}(\omega) = X_{ia}(\omega) + Y_{ia}(\omega)$.

The filtered amplitudes are normalized according to

$$\sum_{ia} |F_{ia}(\omega)|^2 = 1, \tag{S23}$$

and the exciton participation ratio is

$$\mathrm{PR}(\omega) = \frac{1}{\sum_{ia} |F_{ia}(\omega)|^4}. \tag{S24}$$

The corresponding real-space transition density is

$$n_{\mathrm{tr}}(\boldsymbol{r};\omega) = \sum_{ia} F_{ia}(\omega)\phi_i(\boldsymbol{r})\phi_a(\boldsymbol{r}). \tag{S25}$$

The bare transition electrostatic potential is

$$\Phi_{\mathrm{tr}}^{\mathrm{bare}}(\boldsymbol{r};\omega) = \int \frac{n_{\mathrm{tr}}(\boldsymbol{r}';\omega)}{|\boldsymbol{r}-\boldsymbol{r}'|}\, d\boldsymbol{r}'. \tag{S26}$$

Including the nonlocal dielectric response gives the screened transition electrostatic potential,

$$\Phi_{\mathrm{tr}}^{\mathrm{scr}}(\boldsymbol{r};\omega) = \iint \epsilon^{-1}(\boldsymbol{r}-\boldsymbol{r}')\frac{n_{\mathrm{tr}}(\boldsymbol{r}'';\omega)}{|\boldsymbol{r}'-\boldsymbol{r}''|}\, d\boldsymbol{r}'\, d\boldsymbol{r}'', \tag{S27}$$

where the same screened interaction employed in the sBSE Hamiltonian is used consistently in the post-processing analysis.

## 3.7 Spectral validation of all-$\boldsymbol{k}$ and parameterized $v_W$

A comparison between the form-free all-$\boldsymbol{k}$ and parameterized $v_W$ representations is shown in Fig. S3.

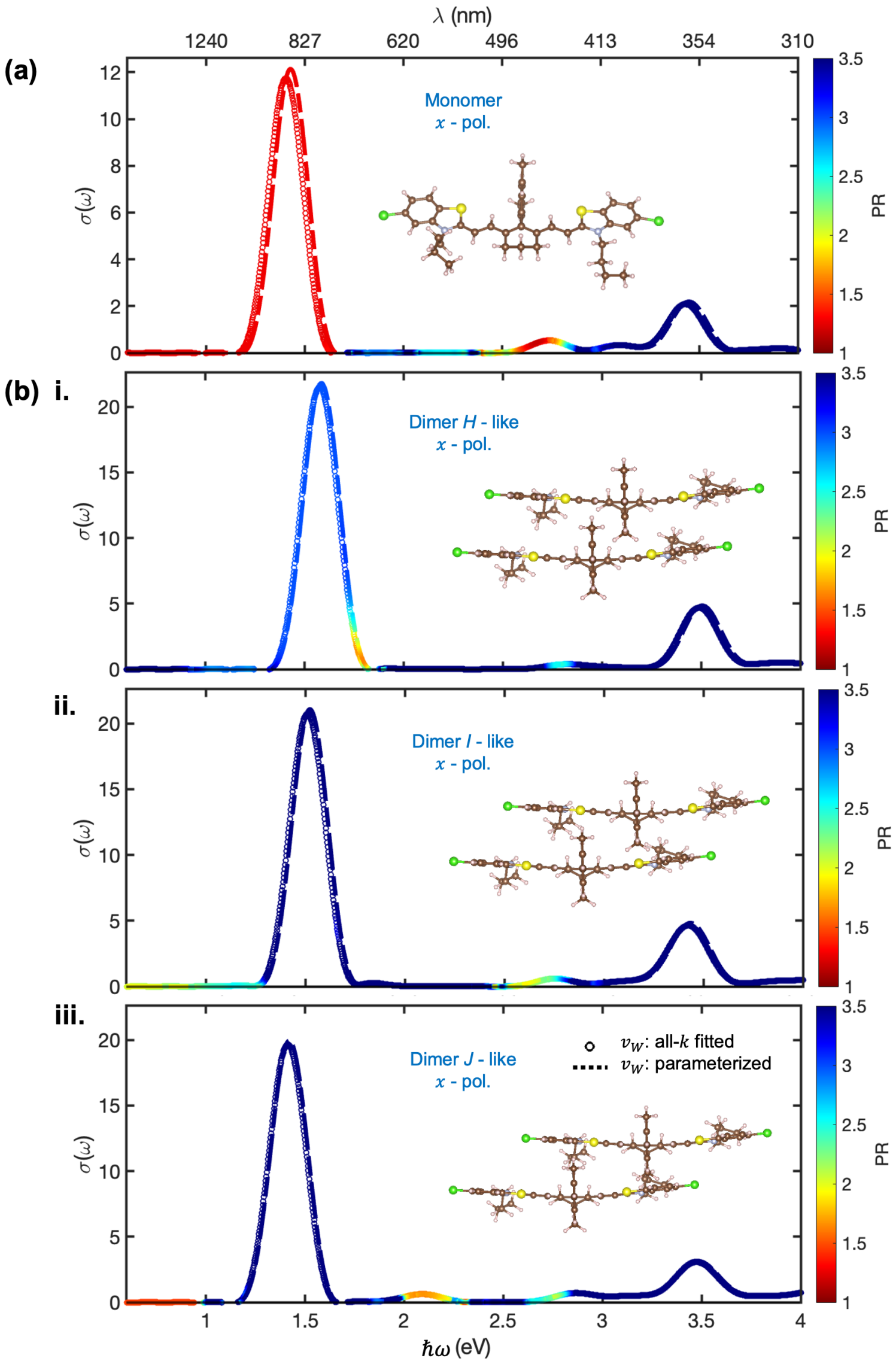


**Fig. S3** **(a)** sBSE absorption spectrum with participation ratio of the Cy7-DiMePh monomer under $x$-polarization. Open-circle spectra correspond to calculations using the all-$\boldsymbol{k}$ fitted screened interaction $v_W$, while dashed curves correspond to a compact parameterized $v_W$.**(b)** Similar to panel **(a)**, but for i. $H$-, ii. $I$-, and iii. $J$-like $2 \times 1$ dimers.

# 4 $v_W$ Parameters

The optimized polynomial-Gaussian coefficients for the monomer and $4 \times 1$ aggregates are listed in Table S2.

**Table S2** Parameterized polynomial-Gaussian expansion coefficients, $\theta_m$, for the parameterized $v_W$ of the Cy7-DiMePh monomer and $4 \times 1$ aggregates.

| $m$ | Monomer | $H$-like $4 \times 1$ | $I$-like $4 \times 1$ | $J$-like $4 \times 1$ |
|---|---|---|---|---|
| 1 | 3.100551 | 4.943122 | 5.021161 | 4.633988 |
| 2 | 23.145005 | 39.898093 | 41.043609 | 36.974514 |
| 3 | -28.500174 | -50.937019 | -52.464969 | -47.125475 |
| 4 | 2.194692 | 5.998689 | 6.314311 | 5.438713 |
| 5 | 44.143239 | 78.298876 | 80.634325 | 73.618073 |
| 6 | 47.928539 | 81.272304 | 83.119569 | 74.474127 |
| 7 | 52.994778 | 88.478807 | 90.022103 | 81.100479 |
| 8 | 25.934359 | 42.135131 | 42.943137 | 38.427649 |
| 9 | 23.697053 | 40.298343 | 41.467467 | 37.823616 |
| 10 | 20.790429 | 36.864283 | 38.078408 | 34.561447 |
| 11 | -9.106951 | -15.244795 | -15.578933 | -13.726605 |
| 12 | -7.700937 | -14.169292 | -14.687103 | -13.388729 |
| 13 | -5.884088 | -12.075102 | -12.625344 | -11.362575 |
| 14 | -17.412537 | -29.954405 | -30.806910 | -27.591424 |
| 15 | -15.961918 | -28.525687 | -29.292327 | -26.128510 |
| 16 | -14.371624 | -26.778774 | -27.859911 | -25.358416 |

# 5 Transition Electrostatic Potential Analysis

To quantify the spatial organization of the screened transition electrostatic potential, the positive and negative potential lobes above a common absolute potential threshold, $\Phi_{\mathrm{cut}}$, are analyzed separately. Their weighted centroids are defined as

$$R_\pm = \frac{\int_{\Omega_\pm} \boldsymbol{r}\, |\Phi_{\mathrm{tr}}^{\mathrm{scr}}(\boldsymbol{r})|\, d\boldsymbol{r}}{\int_{\Omega_\pm} |\Phi_{\mathrm{tr}}^{\mathrm{scr}}(\boldsymbol{r})|\, d\boldsymbol{r}}, \qquad \Omega_\pm = \{\boldsymbol{r} : \pm\Phi_{\mathrm{tr}}^{\mathrm{scr}}(\boldsymbol{r}) \geq \Phi_{\mathrm{cut}}\}, \tag{S28}$$

while the average RMS extent of the positive and negative potential lobes along the polarization direction, i.e., the $x$-axis, is given by

$$L_x = \frac{1}{2} \sum_{\sigma=\pm} \left[ \frac{\int_{\Omega_\sigma} (x - x_\sigma)^2 |\Phi_{\mathrm{tr}}^{\mathrm{scr}}(\boldsymbol{r})| \, d\boldsymbol{r}}{\int_{\Omega_\sigma} |\Phi_{\mathrm{tr}}^{\mathrm{scr}}(\boldsymbol{r})| \, d\boldsymbol{r}} \right]^{1/2}, \tag{S29}$$

where $x_\sigma$ denotes the $x$-coordinate of the corresponding centroid $R_\sigma$.

Quantitative analysis of the screened transition electrostatic potential for the $H$-, $I$-, and $J$-like $4 \times 3$ aggregate clusters is shown in Table S3. The positive and negative potential lobes are analyzed using common absolute potential thresholds ranging from 8 to 16 a.u. At these thresholds, the analyzed isosurfaces are fully enclosed within their native computational boxes for all three aggregate motifs, ensuring that the reported geometric descriptors are free from finite-box truncation effects.

**Table S3** For each $\Phi_{\mathrm{cut}}$, the centroid separation, $d_{\mathrm{c}}$, average longitudinal extent, $L_x$, RMS lobe radius, $R_{\mathrm{RMS}}$, and volume of the high-magnitude potential region, $V_{\mathrm{iso}}$ are computed from the positive and negative regions.

| $\Phi_{\mathrm{cut}}$ (a.u.) | Agg. Motif | $d_{\mathrm{c}}$ (Å) | $L_x$ (Å) | $R_{\mathrm{RMS}}$ (Å) | $V_{\mathrm{iso}}$ (Å$^3$) |
|---|---|---|---|---|---|
| 8 | *H* | 55.38 | 11.88 | 13.23 | 8373 |
| | *I* | 50.38 | 12.43 | 13.58 | 7338 |
| | *J* | 47.33 | 11.09 | 12.35 | 4713 |
| 10 | *H* | 54.63 | 11.40 | 12.45 | 4671 |
| | *I* | 47.47 | 12.44 | 13.28 | 3653 |
| | *J* | 41.30 | 10.10 | 11.15 | 1660 |
| 12 | *H* | 54.86 | 10.82 | 11.65 | 2520 |
| | *I* | 44.52 | 12.18 | 12.83 | 1630 |
| | *J* | 35.23 | 7.52 | 8.67 | 584 |
| 14 | *H* | 56.04 | 10.12 | 10.76 | 1215 |
| | *I* | 37.76 | 10.74 | 11.33 | 566 |
| | *J* | 33.20 | 5.26 | 6.06 | 229 |
| 16 | *H* | 58.27 | 8.51 | 9.02 | 457 |
| | *I* | 29.97 | 8.93 | 9.55 | 174 |
| | *J* | 31.94 | 3.84 | 4.41 | 90 |

# 6 Frenkel Exciton Analysis

## 6.1 Real-space Frenkel projection

Using the site-resolved monomer transition densities defined in the main text, the Frenkel reconstruction is

$$n_{\mathrm{tr}}^{F}(\boldsymbol{r}) = \sum_{i=1}^{N} c_i n_{\mathrm{tr}}^{(i)}(\boldsymbol{r}), \tag{S30}$$

where $N$ is the number of monomers and $\{c_i\}$ are the Frenkel coefficients that satisfy $\sum_i c_i^2 = 1$. The quality of the reconstruction is quantified by the normalized overlap between $n_{\mathrm{tr}}^{F}(\boldsymbol{r})$ and $n_{\mathrm{tr}}^{\mathrm{agg}}(\boldsymbol{r})$, i.e., the projected overlap.[23, 24]

For comparison, an equal-weight Frenkel transition density is also constructed,

$$n_{\mathrm{tr}}^{\mathrm{EW}}(\boldsymbol{r}) = \frac{1}{\sqrt{N}} \sum_{i=1}^{N} n_{\mathrm{tr}}^{(i)}(\boldsymbol{r}), \tag{S31}$$

where each monomer contributes with the same amplitude. The equal-weight overlap is therefore defined as the normalized overlap between $n_{\mathrm{tr}}^{\mathrm{EW}}(\boldsymbol{r})$ and $n_{\mathrm{tr}}^{\mathrm{agg}}(\boldsymbol{r})$.

## 6.2 Optimized Frenkel coefficients

Fig. S4 presents the normalized coefficients obtained from the least-squares projection onto the translated monomer transition-density basis for the $H$-, $I$-, and $J$-like aggregate clusters. Each rectangle represents one monomer in the 12-molecule aggregate, with the color intensity and numerical value indicating its normalized expansion coefficient. The coefficients satisfy the normalization condition, i.e., $\sum_i c_i^2 = 1$.

Although the coefficient distributions differ among the three packing motifs, the transition densities remain well-represented within the Frenkel basis, consistent with the high projection overlaps reported in the main text. These coefficients therefore provide a real-space visualization of how the aggregate excitation is distributed among the localized monomer excitations.

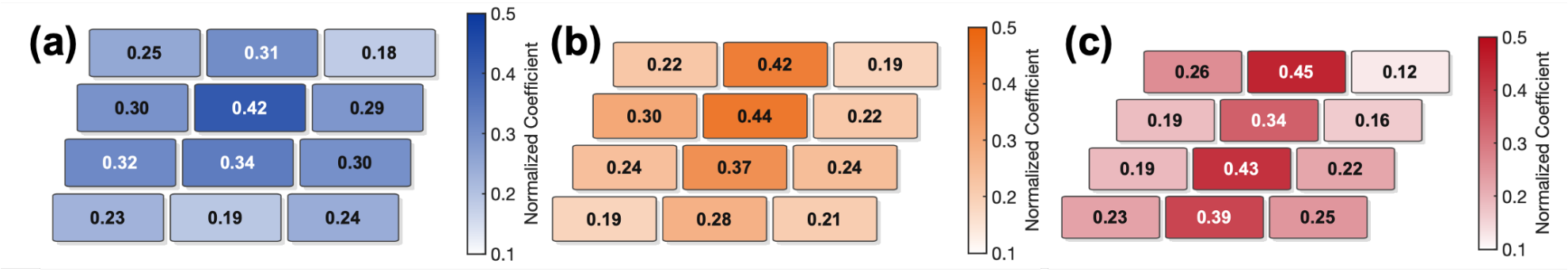


**Fig. S4** Normalized Frenkel expansion coefficients for the **(a)** $H$-, **(b)** $I$-, and **(c)** $J$-like aggregate clusters, obtained by least-squares projection of the sBSE transition density onto the translated monomer transition-density basis. The blue, orange, and red color schemes, respectively, are chosen to indicate the characteristic absorption colors of the three aggregate motifs. Each rectangle represents one monomer in the 12-molecule aggregate, with both the color scale and numerical label indicating the normalized coefficient.

## 6.3 Screening effects in 1D Frenkel excitons

For the transition-charge analysis, the parameterized interaction kernel $v_W(\boldsymbol{k})$ is transformed to its translationally invariant real-space representation, $v_W(\boldsymbol{k}) \xrightarrow{\mathcal{F}^{-1}} v_W(\boldsymbol{r}-\boldsymbol{r}')$. The corresponding real-space inverse dielectric screening factor is

$$\epsilon^{-1}(\boldsymbol{r}-\boldsymbol{r}') = 1 + \frac{v_W(\boldsymbol{r}-\boldsymbol{r}')}{v(\boldsymbol{r}-\boldsymbol{r}')}. \tag{S32}$$

Since the parameterized kernel is translationally invariant, we choose $\boldsymbol{r}' = 0$ as the reference point when evaluating and visualizing the real-space screening function. Thus, the $\epsilon^{-1}(\boldsymbol{r}_{kl})$ used in the screened transition-charge couplings is obtained from the real-space interaction kernel and is not the direct Fourier transform of $\epsilon^{-1}(\boldsymbol{k})$.

Compared with the point-dipole approximation, where the coupling between two selected sites, $n$ and $m$, separated by $r_{nm}$, is screened by $\epsilon^{-1}(r_{nm})$, the screening of individual transition-charge pairs is expected to differ, especially when the dimension of the molecule, or the extent of the transition density, is comparable to or larger than the center-to-center distance. Also, for chromophores with a large aspect ratio, we expect the effect of screening to be more different from the point dipole limit. The transition charges corresponding to the ground and the first excited state of molecules are calculated using the ZINDO method, with chlorine atoms swapped with fluorine.[25, 26]

The unusual observation of "anti-screening" discussed in the main text arises from the combination of the large aspect ratio and strong polarity (extended dipole character) of the transition charge distribution with range-sensitive screening. In the bare coupling case, the contributions from the sum over charge pairs largely cancel themselves, while some of these terms correspond to those between closely separated pairs and are consequently strongly screened, leaving a strong nonzero cumulative contribution. More specifically, at this particular (but highly relevant) slipped geometry, the nearest-neighbor molecules are offset to a degree such that the negative lobe of the transition charge distribution of one molecule lines up with the positive lobe of the other, creating a strong negative contribution to the excitonic coupling. On the other hand, there are the two "diagonal" contributions, referring to the positive-positive and negative-negative charge pairs across the two molecules, each contributing positively at roughly half the amplitude, together with the far-end positive-negative pair, which gives only a negligible negative contribution. These four parts end up nearly canceling one another at their unscreened values. With short-range screening, only the first strong negative contribution is significantly reduced in magnitude, leaving a net positive excitonic coupling between the two molecules.

The above scheme only occurs at a specific value of slip (7.5 Å) for the Cy7-DiMePh dye as expected. To better understand the full picture of the screening effects in this context, we need to explore all plausible packing geometries that are experimentally relevant. In Fig. S5 we examine the dependence of $J_1$ and $J_2$ as functions of the slip offset between successive molecules in a 1D aggregate, and compare the effects of screening. As anticipated, only minute differences can be found between screened and unscreened $J_2$, while $J_1$ is significantly altered. For a constant screening scheme the two curves should be simply related by a multiplicative factor $\epsilon^{-1}(r) = \epsilon_0^{-1}$, where $r \equiv |\boldsymbol{r} - \boldsymbol{r}'|$. Here, instead, the screened $J_1$ is affected not only in terms of magnitude

but also in terms of the onset of sign-flipping as the geometry transitions from an $H$- to a $J$-aggregate.

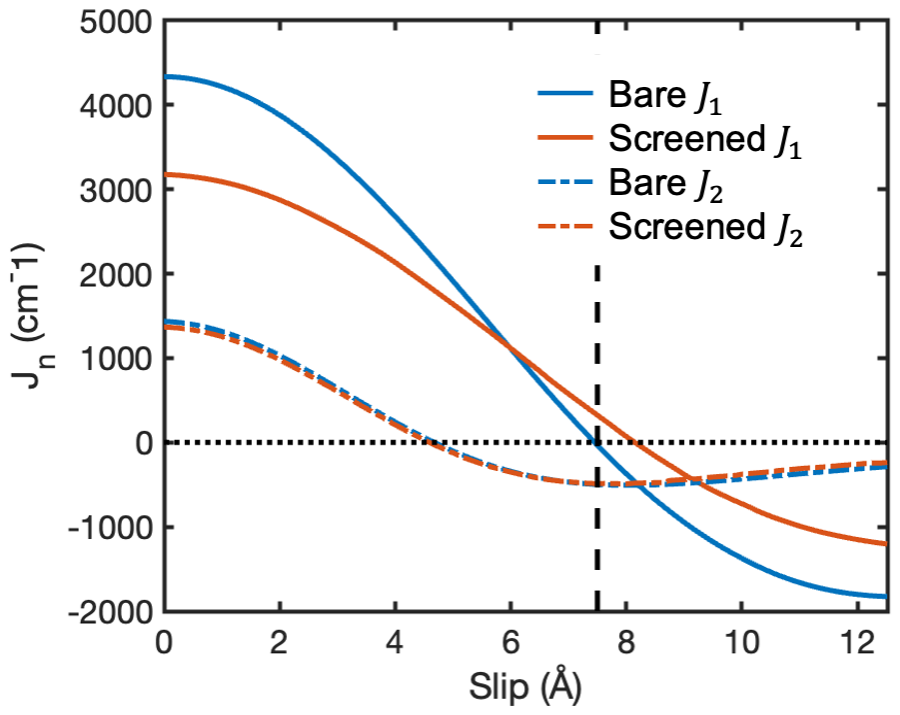


**Fig. S5** The dependence of the nearest-neighbor coupling, $J_1$, on the slip parameter of the 1D aggregate and that of the next-nearest neighbor, $J_2$. Both couplings are shown with and without screening derived from a $J$-like $4 \times 1$ cluster. The 7.5 Å slip used in the main text analysis is indicated by the vertical dashed line.

As a result of the screening-induced change in the excitonic band structure, shown in Fig. S6, the density of states of these systems, while possessing similar van Hove singularity features due to the double-well geometry of the band, exhibits a reversed order of these van Hove peaks, with one of them corresponding to the bright state (vertical dashed lines). Consequently, the unscreened aggregate is predicted to be emissive while the screened one is not.

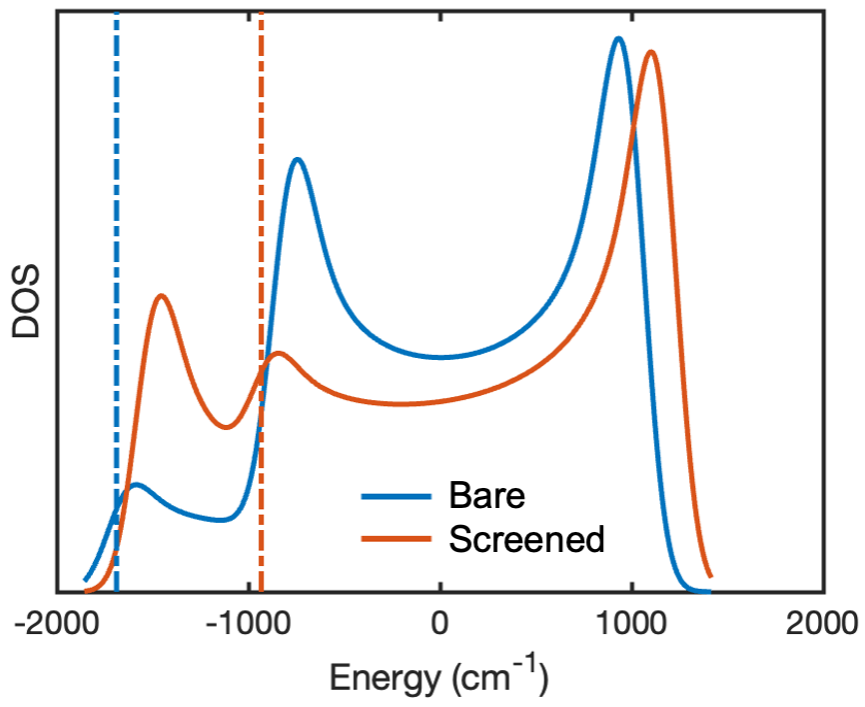


**Fig. S6** Density of states (DOS) corresponding to the unscreened and screened excitonic bands shown in the main text. The screening is derived from a $J$-like $4 \times 1$ cluster.

# 7 Active-space Convergence Check

The active-space convergence of the $4 \times 3$ optical gaps is summarized in Table S4.

**Table S4** Active-space convergence of the $x$-polarized sBSE optical gaps for $4 \times 3$ $H$-, $I$-, and $J$-like aggregates. Linear fits were performed against $1/(N_c + N_v)$ using $E_{\text{gap}} = a/(N_c + N_v) + b$, where $b$ gives the extrapolated infinite-active-space limit.

| $N_v$ | $N_c$ | $H$-like gap (eV) | $I$-like gap (eV) | $J$-like gap (eV) |
|---|---|---|---|---|
| 40 | 120 | 1.72 | 1.68 | 1.52 |
| 80 | 240 | 1.62 | 1.56 | 1.36 |
| 120 | 360 | 1.54 | 1.48 | 1.30 |
| 160 | 480 | 1.50 | 1.42 | 1.24 |
| 200 | 600 | 1.48 | 1.38 | 1.20 |
| 240 | 720 | 1.44 | 1.36 | 1.16 |
| $\infty$ | | 1.42 | 1.33 | 1.13 |
| Difference | | 0.02 | 0.03 | 0.03 |

The difference is computed between the largest active-space calculation, $(N_v, N_c) = (240, 720)$, and the extrapolated infinite-active-space limit.

# 8 Additional Spectral and Electronic Structure Results

Additional spectra for the $3 \times 1$ and $4 \times 1$ aggregates are shown in Fig. S7. Key electronic and optical properties for all systems are collected in Table S5.

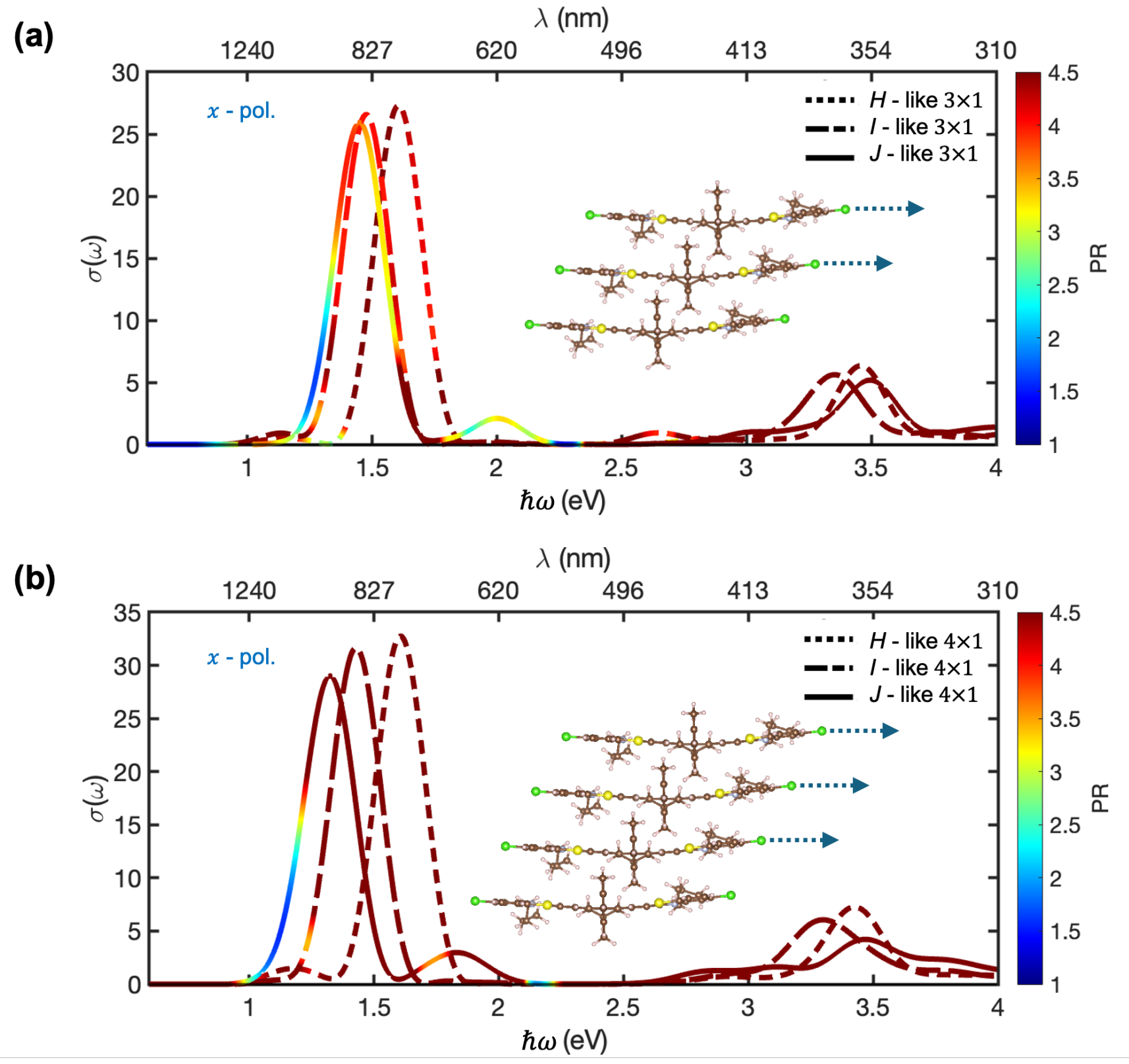


**Fig. S7** sBSE absorption spectra with PRs for 1D Cy7-DiMePh aggregates under $x$-polarization and their individually parameterized $v_W$ kernels. **(a)** $3 \times 1$ $H$-, $I$-, and $J$-like aggregates. **(b)** $4 \times 1$ $H$-, $I$-, and $J$-like aggregates.

**Table S5** Ground- and excited-state properties of the Cy7-DiMePh monomer, 1D aggregates ($2 \times 1$, $3 \times 1$, and $4 \times 1$), and 2D $4 \times 3$ aggregates.

| | **Monomer** | **Dimer $2 \times 1$** | | | **Trimer $3 \times 1$** | | | **Tetramer $4 \times 1$** | | | **Aggregate $4 \times 3$** | | |
|---|---|---|---|---|---|---|---|---|---|---|---|---|---|
| | | *H* | *I* | *J* | *H* | *I* | *J* | *H* | *I* | *J* | *H* | *I* | *J* |
| LDA gap (eV) | 1.102 | 1.061 | 1.058 | 1.027 | 0.834 | 0.817 | 0.782 | 0.591 | 0.551 | 0.521 | 0.442 | 0.379 | 0.370 |
| ngH gap (eV) | 3.687 | 3.592 | 3.607 | 3.625 | 3.386 | 3.283 | 3.330 | 3.101 | 3.063 | 3.048 | 1.409 | 1.143 | 0.795 |
| Bright state (eV) | 1.426 | 1.584 | 1.521 | 1.421 | 1.606 | 1.477 | 1.451 | 1.606 | 1.429 | 1.324 | 1.447 | 1.360 | 1.173 |
| Bright state (nm) | 869 | 782 | 815 | 872 | 772 | 839 | 854 | 772 | 867 | 936 | 856 | 912 | 1057 |
| PR | 1.27 | 2.99 | 3.73 | 5.05 | 4.38 | 4.01 | 3.66 | 6.11 | 6.51 | 5.22 | 15.09 | 13.54 | 12.01 |

LDA and ngH denote the Kohn-Sham and CAM-LDA0 HOMO-LUMO gaps, respectively. Bright-state energies correspond to the most intense $x$-polarized sBSE excitation. PR denotes the participation ratio of the brightest exciton.



\end{document}